\documentclass[aps,pra,twocolumn,showpacs,superscriptaddress]{revtex4} 

\pdfoutput=1
\usepackage{graphicx}  
\usepackage{dcolumn}   
\usepackage{bm}       
\usepackage{amssymb}   
\usepackage{amsmath}

\newcommand{\mean}[1]{{\ensuremath{\big<#1\big>}}}
\newcommand{\comm}[1]{\ensuremath{\big[#1\big]}}

\begin{document}

\title{Generation of directional, coherent matter beams through dynamical instabilities in Bose-Einstein condensates}

\author{Graham R. Dennis}
\author{Mattias T. Johnsson}
\email{mattias.johnsson@anu.edu.au}
\affiliation{Department of Quantum Science, The Australian National University, Canberra, 0200, Australia}
\affiliation{Australian Research Council Centre of Excellence for Quantum-Atom Optics, The Australian National University, Canberra, 0200, Australia}
\date{\today}

\begin{abstract}
We present a theoretical analysis of a coupled, two-state Bose-Einstein condensate with non-equal scattering lengths, and show that dynamical instabilities can be excited. We demonstrate that these instabilities are exponentially amplified resulting in highly-directional, oppositely-propagating, coherent matter beams at specific momenta. To accomplish this we prove that the mean field of our system is periodic, and extend the standard Bogoliubov approach to consider a time-dependent, but cyclic, background. This allows us to use Floquet's theorem to gain analytic insight into such systems, rather than employing the usual Bogoliubov-de Gennes approach, which is usually limited to numerical solutions. We apply our theory to the metastable Helium atom laser experiment of Dall {\emph{et al.}}\ [Phys.\ Rev.\ A \textbf{79}, 011601(R) (2009)] and show it explains the anomalous beam profiles they observed. Finally we demonstrate the paired particle beams will be EPR-entangled on formation.
\end{abstract}

\maketitle 

\section{Introduction}

The experimental realization of Bose-Einstein condensates (BECs) \cite{andersonET1995,davisET1995} has allowed the creation of macroscopic quantum systems that are sensitive, isolated from the environment, and highly controllable, providing an excellent system in which to investigate many body quantum mechanics \cite{dalfovoET1999,greinerET2002,blochET2008,pollackET2009}, as well as leading to the creation of the atom laser, a matter-wave analog of the optical laser \cite{mewesET1997,hagleyET1999,ballaghET2000}. 

Atom lasers are usually generated by creating a BEC in which the atoms are trapped by magnetic or optical fields, and then changing the state of a subset of the atoms so that they are no longer trapped, and are free to escape. This can be done via RF spin flipping \cite{mewesET1997}, or Raman outcoupling \cite{hagleyET1999,robinsET2006}, respectively resulting in a beam that falls under gravity, or a beam that can have some directionality due to a momentum transfer corresponding to a few cm$\,$s$^{-1}$. Such coherent, directional matter beams offer great promise not only as a sensitive probe of the source condensate, but also offer potential in terms of atom interferometry and precision sensors. Furthermore, both BECs and atom lasers also offer the possibility of creating non-classical states that exhibit properties such as number and quadrature squeezing \cite{chuuET2005,Johnsson2007,liET2008,esteveET2008,Haine2009}, which among other things have the potential to enhance the sensitivity of atom interferometers \cite{dowling1998}, allowing the measurement of electric, magnetic and gravitational fields, as well as rotations, with great precision \cite{toriiET2000,wichtET2001,cladeET2006,lecoqET2006}. 

The creation of coherent matter beams can also be achieved by utilizing the inherent properties of the BEC itself. This can be more useful than a conventional atom laser as not only can such beams carry information about the interaction properties, dynamics and quantum state of the BEC, but depending on their method of generation they can possess highly interesting quantum states of their own. The dissociation of a molecular BEC has been shown, for example, to lead to the creation of pairs of particles with oppositely-directed momenta that are expected to exhibit EPR correlations \cite{Kheruntsyan2005}, as can different forms of atomic four-wave mixing \cite{Hilligsoe2005,Ferris2009,perrinET2008,ogren2007}. One can also utilize colliding condensates resulting in a spherical shell of scattered atoms where opposite points on the sphere have correlations that depend on the the interation properties of the atoms within the condensate. Most of these schemes, however, result in atomic fields that are not in easily accessible modes. They rely, for example, on collisions or dissociation resulting in scattering over the full sphere, beams that are co-propagating rather than spatially distinct, or that have their interesting quantum properties such as entanglement spread over a large number of resonant modes.

In this paper we present an analysis of a new method of generating correlated atomic fields, which has the advantages of being experimentally realistic and producing spatially distinct, coherent, highly-directed atomic beams at well-defined momenta. The beams are produced due to dynamic instabilities within the atomic system itself, and as such do not rely on molecular dissociation or colliding condensates, or the existence of engineered quantum optical fields to use as a template for the desired quantum statistics to be transferred to the atomic fields.

Our scheme is motivated by the experiment of Dall {\emph{et al.}}\ \cite{Dall2009}, who exploited the combination of a metastable helium BEC and a double stacked multichannel plate to image the transverse (i.e. orthogonal to the propagation direction) beam profile of an atom laser. At high outcoupling powers they observed an anomalous particle flux that was ejected along the long axis of the BEC, and was well separated from the bulk atom laser profile. In this paper we develop a general theoretical model of a class of systems which includes that experiment, and demonstrate that pairs of correlated beams at specific quantized momenta can be produced.

In brief, our scheme requires a two-state BEC that has a high aspect ratio, with the two states being coupled in some way. A further requirement is that the $s$-wave scattering length of the atoms within each state and between the states must not all be equal. One experimental realization of such a system would be a magnetically trapped condensate with two equally-trapped Zeeman states, for example the $F=2$, $m_F=1$ and $F=1$, $m_F=-1$ levels of $^{87}\text{Rb}$.  Another would be an optically trapped condensate using the quadratic Zeeman shift to lift the degeneracy in the Zeeman level splittings to create an effective two-level system.  In both the optical and magnetically trapped cases the two levels could be coupled by two optical fields utilizing a Raman transition.  Two-state Raman coupling has been achieved experimentally in $^{87}\text{Rb}$ \cite{Debs2009}.

The existence of the coupling between the two levels, as well as the non-equal scattering lengths, results in dynamic instabilities in the condensate which can grow exponentially at specific well-defined momenta. As we will demonstrate, the equations governing the unstable system can be mapped to those describing optical parametric down-conversion, and have the same result --- the generation of pairs of entangled particles with oppositely directed momenta. Only the unstable momentum modes which have de Broglie wavelengths that fit within the condensate are amplified, which results in the pairs being emitted along the long (weakly trapped) axis of the condensate. If the aspect ratio of the condensate is sufficiently large, a pair of highly directed beams will be generated.

As part of our analysis, we will develop techniques for the application of Bogoliubov theory to perturbations about a time-dependent mean field background, rather than a static one. A Bogoliubov approach is necessary because the creation of the entangled pairs of particles is spontaneously seeded, meaning that analysis beyond the mean field is required. In its standard form Bogoliubov theory assumes a time-independent mean field. In our case, however, the effect depends on the existence of a \emph{time-dependent} mean field background, which suggests a Bogoliubov-de Gennes approach should be used. The usual drawback to such an approach is that the Bogoliubov-de Gennes equations are generally analytically intractable and are therefore solved numerically, which in many cases precludes deep insight into the problem. In this paper, however, we prove that the mean field background in our system, while time-dependent, is also periodic. This enables us to employ Floquet's theorem to gain a semi-analytic solution to the problem of determining the excitation spectrum and instabilities of the system.

The outline of this paper is follows: We begin by giving a brief overview of Bogoliubov theory in Section~\ref{secBogOverview}, and demonstrate that the basic approach is still sound even in the presence of a time-dependent background. In Section~\ref{secPeriodicSolutions} we solve for the mean field dynamics of our system, and map them to modified optical Bloch equations with a nonlinear term. This enables us to prove that the mean fields, while time-dependent, are also periodic, regardless of the initial state of the system. The equations of motion for the fluctuations beyond the mean field are derived in Section~\ref{SectionExcitationDynamics}, and we show that in the special case of equal nonlinearities for the two atomic states they reduce to the standard Bogoliubov excitation spectrum. In Section~\ref{secInstabilitiesAndExcitations} we take advantage of the temporal periodicity of the mean field to employ Floquet's theorem and show that the Floquet exponents can indicate dynamical instabilities in the system.  These dynamical instabilities lead to exponential growth of excitations at particular momenta. We also check the accuracy of the theory by performing truncated Wigner simulations of the system to ensure the excitations occur where our Floquet theory predicts, and derive the equations of motion for the annihilation and creation operators governing the excitations. As a further test of the validity of our theory we apply it to the experiment of Dall \emph{et al.}\ \cite{Dall2009} in Section~\ref{secTheoryExperimentAgreement} and demonstrate that the anomalous particle production they measured is correctly described by our theory. In Section~\ref{secEntanglement} we note that the equations of motion for the excitations are identical to those governing optical parametric down conversion, and thus should exhibit EPR entanglement, at least on formation. Finally, in Section~\ref{secDiscussion}, we discuss the optimal experimental system to generate the dynamical instabilities our theory predicts, and consider possible difficulties, and conclude in Section~\ref{secConclusion}.

\section{Overview of Bogoliubov theory}
\label{secBogOverview}

As described in the Introduction, the paired matter beams generated in our scheme arise from a dynamical instability in the condensate, and in order to determine which modes are dynamically unstable, the excitation spectrum of the condensate must be obtained. To that end, we begin with an overview of the theory of condensate excitations beyond the mean field approximation, and then use these results to calculate the stability of the condensate and its excitation spectrum in Section \ref{SectionExcitationDynamics}. More comprehensive treatments of the Bogoliubov theory can be found in \cite{PethickSmith2002} and in a number of review articles (see, for example, \cite{Legett2001, Ozeri2005, Proukakis2008}).

The problem is to determine the response of the condensate to small fluctuations about the mean field. Typically the condensate is stable to such fluctuations, and their energy spectrum determines the phase and group velocities of the excitations.  In the case that the condensate is dynamically unstable some modes will undergo exponential growth, which corresponds to the generalised energy spectrum containing non-zero imaginary components. As a concrete example of the techniques we will employ, consider a single-component condensate described by the Hamiltonian
\begin{align}
    \label{eqElementaryExcitationsExampleHamiltonian}
    \hat{H} &= \int d \mathbf{x}\, \hat{\Psi}^\dagger \left( -\frac{\hbar^2 \nabla^2}{2 M} + V(\bm{x}) + \frac{1}{2} U \hat{\Psi}^\dagger \hat{\Psi} - \mu \right) \hat{\Psi},
\end{align}
where an arbitrary energy offset $\mu$ has been included. This term is introduced for calculational reasons and has no physical influence on the Hamiltonian as it cannot affect any expectation values. This can be seen by noting that although it contributes a different energy offset for states with different total number, these states are uncoupled due to the Hamiltonian (\ref{eqElementaryExcitationsExampleHamiltonian}) conserving particle number. For the case of
calculating the excitation spectrum of a ground state, the arbitrary energy offset $\mu$ will be identified as the condensate chemical potential.

To find the excitation spectrum of Eq.~(\ref{eqElementaryExcitationsExampleHamiltonian}) we use the Gross-Pitaevskii equation to describe the mean field of the condensate, and then consider quantum-mechanical fluctuations about this mean field. To this end we define the deviation operator $\delta \hat{\Psi} = \hat{\Psi} - \Psi$, where $\Psi = \mean{\hat{\Psi}}$, and $\delta \hat{\Psi}$ is considered to be a small quantity in the sense that its quadratic moments are small compared to $\left|{\Psi}\right|^2$ (its first-order moments are identically zero).

The concept of $\delta \hat{\Psi}$ being a ``small quantity" is not entirely trivial. For it to be considered in any sense small, the mean field $\mean{\hat{\Psi}}$ must be non-zero. However the state of condensates with a large number of atoms is well approximated by either a state with well-defined total number or as a statistical mixture of coherent states with random phases \cite{Leggett2001}. In both of these cases the mean field $\mean{\hat{\Psi}}$ is zero, as the system has no well-defined global phase. As any \emph{physical} expectation value is independent of the choice of global phase, however, any analysis can be performed for a coherent state with a specific global phase, and the results then averaged over the global phase. Because all observables will be made up of equal numbers of annihilation and creation operators (since atom number is conserved), the choice of global phase cannot affect the final result.

An alternative justification is to note that spontaneous symmetry breaking results in the selection of a specific phase, which is equivalent to assuming that the state of the condensate is unchanged on the addition of a particle, that is the states $|N\rangle$ and $|N+1\rangle$ are physically identical \cite{LeggettET1991,PitaevskiiET2003}. This is a robust assumption for BECs with particle number much greater than one, which is the case for most bulk condensate experiments.


Making the replacement $\hat{\Psi} \rightarrow \Psi + \delta \hat{\Psi}$, the Hamiltonian Eq.~(\ref{eqElementaryExcitationsExampleHamiltonian}) can be expanded in powers of $\delta \hat{\Psi}$,
\begin{align}
    \label{eqHamiltonianPowerSeriesExpansion}
    \hat{H} &= \hat{H}_0 + \hat{H}_1 + \hat{H}_2 + \hat{H}_3 + \hat{H}_4,
\end{align}
where $\hat{H}_n$ contains terms of order $(\delta\hat{\Psi})^n$. The excitation spectrum of the Hamiltonian $\hat{H}$ is then approximately given by the eigenvalue spectrum of the lowest order non-trivial term in Eq.~(\ref{eqHamiltonianPowerSeriesExpansion}).

The zeroth order term in Eq.~(\ref{eqHamiltonianPowerSeriesExpansion}),
\begin{align}
    \hat{H}_0 &= \int d \mathbf{x}\, \Psi^* \left( -\frac{\hbar^2 \nabla^2}{2M} + V(\mathbf{x}) + \frac{1}{2} U \big|\Psi\big|^2 - \mu \right) \Psi,
\end{align}
is simply a constant and represents the total energy of the unexcited mean field. The first order term is of the form
\begin{align}
    \hat{H}_1 &= \int d \mathbf{x}\, \delta \hat{\Psi}^\dagger \left(i \hbar \frac{\partial \Psi}{\partial t} \right)  + \int d \mathbf{x}\, \left(i \hbar \frac{\partial \Psi}{\partial t} \right)^* \delta \hat{\Psi},
\end{align}
where the mean field $\Psi$ evolves as
\begin{align}
    i \hbar \frac{\partial\Psi}{\partial t} &= \left(-\frac{\hbar^2 \nabla^2}{2 M} + V(\mathbf{x}) + U \big| \Psi\big|^2 - \mu \right) \Psi.
\end{align}

Although the first order term $\hat{H}_1$ is non-zero, it does not affect the evolution of the deviation operator. To see this we note that as $\hat{\Psi} = \Psi + \delta \hat{\Psi}$ we have
\begin{equation}
    i \hbar \frac{\partial \Psi}{\partial t} + i \hbar \frac{\partial }{\partial t}\delta{\hat{\Psi}} = \comm{\Psi, \hat{H}} + \comm{\delta\hat{\Psi}, \hat{H}}
\end{equation}
and because $\Psi$ is a complex-valued function it commutes with all operators, giving
\begin{align}
    i \hbar \frac{\partial }{\partial t}\delta \hat{\Psi} &= \comm{\delta \hat{\Psi}, \hat{H}} - i \hbar \frac{\partial  \Psi}{\partial t} \nonumber \\
    &= \comm{\delta \hat{\Psi}, \hat{H}_1} + \comm{\delta \hat{\Psi}, \hat{H}_2 + \hat{H}_3 + \hat{H}_4} - i \hbar \frac{\partial\Psi }{\partial t} \nonumber\\
    &= i \hbar \frac{\partial \Psi}{\partial t} + \comm{\delta \hat{\Psi}, \hat{H}_2 + \hat{H}_3 + \hat{H}_4} - i \hbar \frac{\partial \Psi}{\partial t} \nonumber\\
    &= \comm{\delta \hat{\Psi}, \hat{H}_2 + \hat{H}_3 + \hat{H}_4}. 
\label{eqDeviationOperatorEvolution}
\end{align}
Since $\hat{H}_1$ does not occur in Eq.~(\ref{eqDeviationOperatorEvolution}), it does not affect the evolution of the deviation operators and will not contribute to the excitation spectrum. We note that typical treatments of the Bogoliubov theory consider the restricted case of a static condensate density and choose $\mu$ as the chemical potential such that $\partial \Psi / \partial t = 0$, and hence $\hat{H}_1=0$. As shown by Eq.~(\ref{eqDeviationOperatorEvolution}), this choice of $\mu$ is unnecessary as $\hat{H}_1$ does not influence the evolution of $\delta\hat{\Psi}$, \emph{independent} of the choice of $\mu$ and even in the general case of a non-stationary mean field. This is relevant as we will be considering the extension of this theory to a condensate with a time-dependent density background.

The first physically important term in Eq.~(\ref{eqHamiltonianPowerSeriesExpansion}) is
\begin{align}
    \begin{split}
        \hat{H}_2 &= \int d \mathbf{x}\, \delta \hat{\Psi}^\dagger \left(-\frac{\hbar^2 \nabla^2}{2 M} + V(\mathbf{x}) + 2 U\big|\Psi\big|^2 -\mu \right)\delta\hat{\Psi}\\
         &\phantom{=} + \frac{1}{2} U\int d \mathbf{x}\, \left(\Psi^2 \delta \hat{\Psi}^\dagger\delta \hat{\Psi}^\dagger  +  (\Psi^*)^2 \delta \hat{\Psi} \,\delta \hat{\Psi}\right).
    \end{split}
    \label{eqHamiltonianPowerSeriesQuadraticTerm}
\end{align}
As the deviation operator $\delta \hat{\Psi}$ is small compared to the mean field $\Psi$, the higher-order contributions $\hat{H}_3$ and $\hat{H}_4$ to the total Hamiltonian can then be neglected compared to $\hat{H}_2$. The excitation spectrum of the condensate about the mean field $\Psi$ is then approximately given by the energy spectrum of $\hat{H}_2$.

To avoid solving the infinite dimensional eigenvalue problem $\hat{H}_2 |\Psi \rangle = E |\Psi \rangle$ for the condensate excitation spectrum directly,  it is desirable to apply a linear transformation to $\hat{H}_2$ that will diagonalize it in the form
\begin{align}
    \label{eqQuadraticHamiltonianAnsatz}
    \hat{H}_2 &= \sum_i \hbar \omega_i \hat{\Lambda}_i^\dagger \hat{\Lambda}_i^{\phantom{\dagger}},
\end{align}
for some boson annihilation operators $\hat{\Lambda}_i$ and real frequencies $\omega_i$. In this form, the Hamiltonian can be simply interpreted as representing a set of modes with energies $\hbar \omega_i$, which is the condensate excitation spectrum. In general, it is not possible to transform $\hat{H}_2$ into the form Eq.~(\ref{eqQuadraticHamiltonianAnsatz}) if the Hamiltonian possesses any instabilities \cite{Leonhardt2003}, however one frequently considers the excitation spectrum of the ground state which is stable by definition and in this case such a transformation is always possible.

In the general case, we look for the operators $\hat{\Lambda}_i$ satisfying
\begin{align}
    \label{eqElementaryExcitationsEvolution}
    i \hbar \frac{\partial }{\partial t}\hat{\Lambda}_i &= \comm{\hat{\Lambda}_i, \hat{H}_2 } = - \hbar \omega_i \hat{\Lambda}_i
\end{align}
where $\omega_i$ is real if and only if $\hat{\Lambda}_i$ is a boson annihilation operator \cite{Leonhardt2003}. In the case that $\omega_i$ is complex, boson annihilation operators can be constructed from the $\hat{\Lambda}_i$ as discussed in Appendix \ref{appExcitationEvolution}. Hence $\hbar\omega_i$ can be considered to be a generalised energy spectrum of the condensate where non-zero imaginary components correspond to dynamical instabilities of the condensate. Note that although the eigenvalues of $\hat{H}_2$ must be real as it is Hermitian, the eigenvalues of Eq.~(\ref{eqElementaryExcitationsEvolution}) need not be real. For example, the Hamiltonian for degenerate parametric down-conversion $\hat{H} = \hbar\chi \left(\hat{a}\hat{a} + \hat{a}^\dagger \hat{a}^\dagger \right)$ is Hermitian but the corresponding eigenvalues of Eq.~(\ref{eqElementaryExcitationsEvolution}) are pure imaginary. In this case the occupation of the mode $\hat{a}$ undergoes exponential growth. The case of complex eigenvalues $\omega_i$ is discussed further in Sections \ref{SectionExcitationDynamics} and \ref{secInstabilitiesAndExcitations}, and Appendix \ref{appExcitationEvolution}.

Equation (\ref{eqElementaryExcitationsEvolution}) is most easily solved by expanding the $\hat{\Lambda}_i$ in a complete, linearly independent basis $\{\hat{\Upsilon}_j\}$ such that $\hat{\Lambda}_i = \bm{c}_i^\dagger \hat{\bm{\Upsilon}}$ where $\bm{c}_i$ is a complex vector, $\bm{c}_i^\dagger$ denotes its conjugate-transpose, and $\hat{\bm{\Upsilon}}$ is the column vector formed by the complete basis $\{\hat{\Upsilon}_j\}$. For the case of Eq.~(\ref{eqHamiltonianPowerSeriesQuadraticTerm}), an appropriate basis is $\{\hat{\Upsilon}_j\} = \{\delta\hat{\Psi}, \delta\hat{\Psi}^\dagger\}$. As the Hamiltonian $\hat{H}_2$ is quadratic, its commutator with every operator $\hat{\Upsilon}_j$ will be linear in the operators $\{\hat{\Upsilon}_j\}$. Defining the complex matrix $\mathcal{H}$ to represent this relationship
\begin{align}
    \label{eqScriptHRelationshipToHamiltonian}
    \sum_k \mathcal{H}_{jk} \hat{\Upsilon}_k &= \comm{\hat{\Upsilon}_j, \hat{H}_2}
\end{align}
permits Eq.~(\ref{eqElementaryExcitationsEvolution}) to be recast as an eigenvalue problem in $\mathcal{H}$,
\begin{align}
    \comm{\bm{c}_i^\dagger \hat{\bm{\Upsilon}}, \hat{H}_2} &= \bm{c}_i^\dagger \mathcal{H} \hat{\bm{\Upsilon}} = - \hbar \omega_i \bm{c}_i^\dagger \hat{\bm{\Upsilon}}, \label{eqElementaryExcitationsEigenvalueProblemWithOperators}\\
    \implies \bm{c}_i^\dagger \mathcal{H} &= - \hbar \omega_i \bm{c}_i^\dagger \label{eqElementaryExcitationsEigenvalueProblem}
\end{align}
where the last line follows as the components of $\hat{\bm{\Upsilon}}$ are linearly independent. If the mean field $\Psi$ is time-independent, then the matrix $\mathcal{H}$ will also be time-independent and Eq.~(\ref{eqElementaryExcitationsEigenvalueProblem}) represents an eigenvalue problem for the left eigenvectors $\bm{c}_i^\dagger$ and eigenvalues $-\hbar \omega_i$ of the matrix $\mathcal{H}$. If the mean field $\Psi$ simply evolves due to a global phase rotation, this can be cancelled by appropriate choice of the arbitrary energy offset $\mu$ making $\mathcal{H}$ time-independent.  In the case of the condensate ground state, that offset will be the chemical potential of the condensate. The eigenvalues $\{\hbar \omega_i\}$ then represent the generalised excitation spectrum of the condensate about the mean field, which was to be determined. 

It is important to note that for the eigenvalues of $\mathcal{H}$ to determine the solution to Eq.~(\ref{eqElementaryExcitationsEvolution}) the matrix $\mathcal{H}$ must be constant. In the next section these techniques will be generalised to the case of a \emph{periodic} mean field in which the time-dependence of the matrix $\mathcal{H}$ cannot be removed by any analytic transformation.

In the case of a homogeneous condensate ($V(\mathbf{x}) = 0$) the matrix $\mathcal{H}$ can be diagonalized analytically to give the condensate excitation spectrum as
\begin{align}
    \hbar \omega(\mathbf{k}) &= \sqrt{\varepsilon(\mathbf{k})\left(\varepsilon(\mathbf{k}) + 2 n U \right)},
    \label{eqBogoliubovSpectrum}
\end{align}
where $\mathbf{k}$ is the wave vector, $\varepsilon(\mathbf{k}) = \hbar^2 \mathbf{k}^2 / 2 M$ is the free-particle energy spectrum and $n = \big|\Psi \big|^2$ is the condensate density. Equation~(\ref{eqBogoliubovSpectrum}) is known as the Bogoliubov excitation spectrum \cite{Bogoliubov1947}.  

The interested reader is referred to the review paper by Ozeri for further details about Bogoliubov theory in Bose-Einstein condensates \cite{Ozeri2005}.

\section{Mean field dynamics and existence of periodic solutions}
\label{secPeriodicSolutions}




As described in the Introduction, our scheme consists of a two-level condensate with the two levels $|0\rangle$ and $|1\rangle$ coupled in some way, leading to Rabi oscillations between the levels.




Under this approximation, the second-quantized Hamiltonian for the system is given by
\begin{align}
    \label{eqInitialHamiltonian}
    \begin{split}
    \hat{H} &= \sum_{i=1,0} \int d\mathbf{x}\, \hat{\Psi}_i^\dagger \left(\frac{-\hbar^2 \nabla^2}{2 M} - \mu\right)\hat{\Psi}_i^{\phantom{\dagger}} \\
            & + \frac{1}{2} \sum_{i j = 1, 0} U_{i j}\int d\mathbf{x}\, \hat{\Psi}_i^\dagger \hat{\Psi}_j^\dagger \hat{\Psi}_j^{\phantom{\dagger}} \hat{\Psi}_i^{\phantom{\dagger}}  \\
            &\phantom{=} + \hbar \Omega \int d\mathbf{x}\, \left(\hat{\Psi}_1^\dagger \hat{\Psi}_0^{\phantom{\dagger}} + \hat{\Psi}_0^\dagger \hat{\Psi}_1^{\phantom{\dagger}}\right),
    \end{split}
\end{align}
where $U_{ij} = 4\pi \hbar^2 a_{ij}/M$ is the nonlinear interaction strength, $a_{ij}$ is the s-wave scattering length between internal states $|i\rangle$ and $|j\rangle$, $\Omega$ is the Rabi frequency which is taken to be real, and $\mu$ is an energy offset which has been included to cancel the global phase rotation which would otherwise be present. The equations of motion corresponding to this Hamiltonian are
\begin{eqnarray}
i \hbar \frac{\partial }{\partial t}\hat{\Psi}_1  = -\frac{\hbar^2}{2M}\nabla^2 \hat{\Psi}_1  &+ U \left(\hat{\Psi}_1^\dagger \hat{\Psi}_1^{\phantom{\dagger}} + \hat{\Psi}_0^\dagger \hat{\Psi}_0^{\phantom{\dagger}}\right) \hat{\Psi}_1 \nonumber \\
&+ \hbar \Omega \hat{\Psi}_0 - \mu \hat{\Psi}_1,  \\
i \hbar \frac{\partial }{\partial t}\hat{\Psi}_0 = -\frac{\hbar^2}{2M} \nabla^2 \hat{\Psi}_0 &+ U \left(\hat{\Psi}_1^\dagger \hat{\Psi}_1^{\phantom{\dagger}} + \kappa \hat{\Psi}_0^\dagger \hat{\Psi}_0^{\phantom{\dagger}} \right) \hat{\Psi}_0 \nonumber \\
&+ \hbar \Omega \hat{\Psi}_1 - \mu \hat{\Psi}_0,
\label{eqOperatorEquationsOfMotion}
\end{eqnarray}
where $U=U_{11}=U_{10}$ and we have defined the nonlinearity mismatch parameter
\begin{equation}
\kappa = U_{00}/U_{11}.
\label{eqKappaDefinition}
\end{equation} 

As described in Section~\ref{secBogOverview}, the excitation spectrum of a condensate can be obtained by approximating each field operator as a complex number (the mean field) plus a small fluctuation term, and then either diagonalizing the Hamiltonian \cite{Bogoliubov1947,FetterWalecka} or diagonalizing the linearized equations of motion for the fluctuations themselves. We will take the latter approach in this paper, but with the difference that the mean field about which the linearization procedure will take place is itself \emph{time-dependent}.


The method for diagonalizing the evolution equations of the linearized fluctuations to obtain the excitation spectrum about the ground state is the same method used to determine the stability of fixed points of systems of nonlinear ordinary differential equations. As mentioned in the previous section, this method relies critically on the fact that it is a stationary solution about which the equations are linearized. Floquet's Theorem \cite{AppliedNonlinearDynamics} allows the stability of \emph{periodic} solutions to be considered, and it is this theorem that will be used to determine the stability of the condensate about these mean field dynamics. We will now show that the mean field evolution of Eq.~(\ref{eqOperatorEquationsOfMotion}) is periodic.

\begin{figure}
    \centering
    \includegraphics[width=9cm]{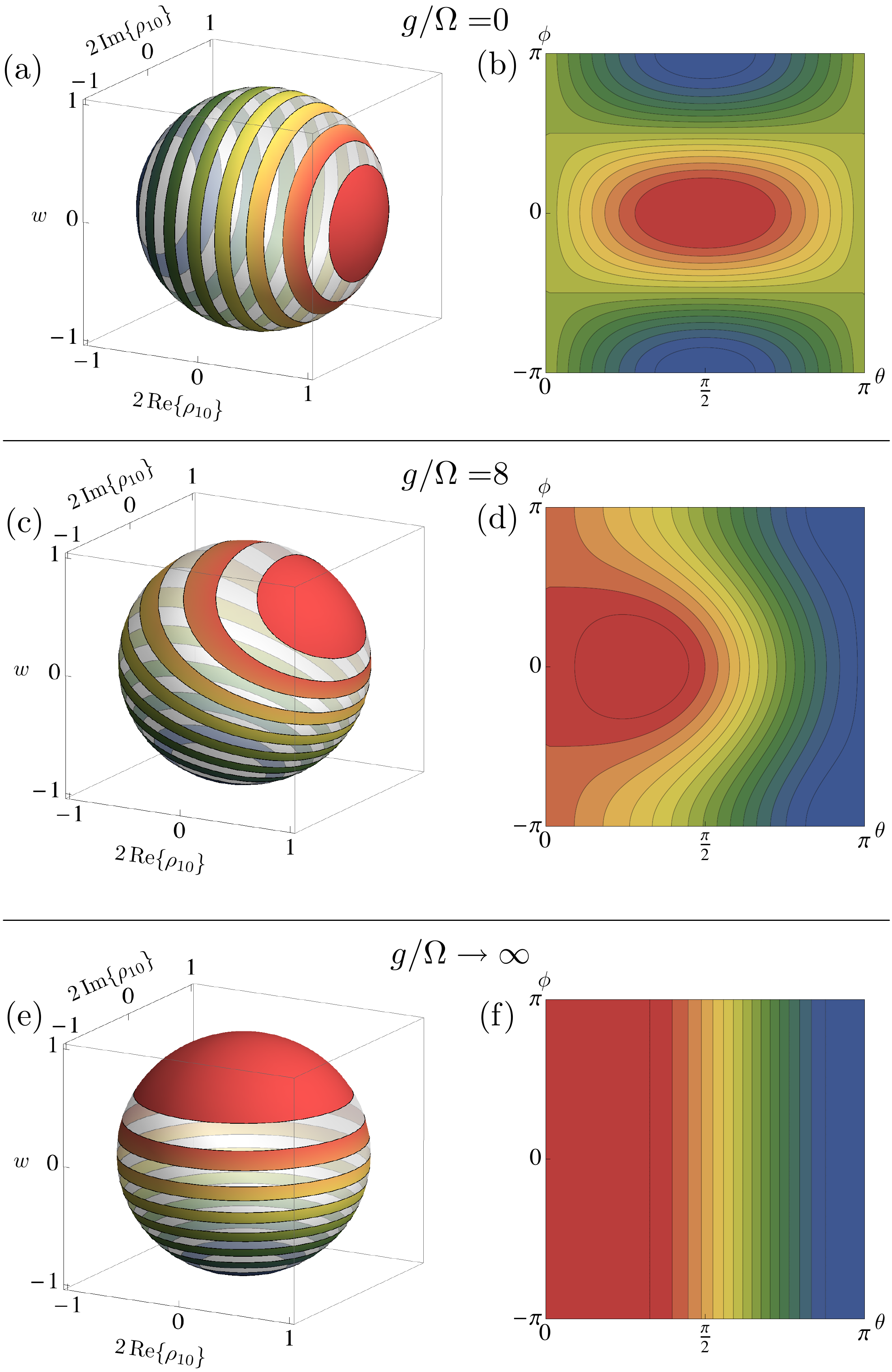}
    \caption{(color online) Bloch sphere representation of the evolution described by Eq.~(\ref{eqOpticalBlochEquations}). The upper figures (a) and (b) represent the case of the usual optical Bloch equations with no damping ($g/\Omega=0$), middle figures (c) and (d) illustrate the effect of the nonlinear term on the evolution with $g/\Omega = 8$, and the lower figures (e) and (f) illustrate the limit where the nonlinear term dominates ($g/\Omega \rightarrow \infty$). The left figures (a), (c), (e) illustrate the Bloch sphere colored according to the energy per particle $E$ (see Eq.~(\ref{eqOpticalBlochEnergy})). The system is constrained to move on lines of constant color. Bands have been removed from these spheres for illustration purposes only. The right figures (b), (d), (f) are contour plots of the energy per particle $E$ over the surface of the Bloch sphere.
     \label{figBlochSphere}}
\end{figure}

Within the local density approximation \cite{Stamper-Kurn1999,Zambelli2000}, we can assume that the mean field remains homogeneous; only the excitations will have spatial dependence. The equations of motion for the mean field then reduce to the following ordinary differential equations,
\begin{subequations}
    \label{eqMeanFieldEquationsOfMotion}
    \begin{align}
    i \hbar \frac{d }{dt}\Psi_1 &= U \left( |\Psi_1|^2 + |\Psi_0|^2 \right)\Psi_1 + \hbar \Omega \Psi_0 - \mu \Psi_1,\\
    i \hbar \frac{d }{dt}\Psi_0 &= U \left( |\Psi_1|^2 + \kappa |\Psi_0|^2 \right)\Psi_0 + \hbar \Omega \Psi_1 - \mu \Psi_0.
    \end{align}
\end{subequations}
Although solving Eq.~(\ref{eqMeanFieldEquationsOfMotion}) for $\kappa \neq 1$ is intractable analytically, it can at least be shown that the solutions are periodic up to a global phase rotation, and exactly periodic for appropriate choice of the arbitrary energy offset $\mu$. The simplest way to do this is to recognise that these equations are modified optical Bloch equations containing a nonlinear term but with no damping. Defining $\Psi_i = c_i\sqrt{n}$ where $n = |\Psi_1|^2 + |\Psi_0|^2$ is the total density, the equations of motion for the density matrix terms $\rho_{10} = c_{1}^{}c_{0}^*$ and $w = \rho_{11}-\rho_{00} = |c_1|^2 - |c_0|^2$ are
\begin{subequations}
    \label{eqOpticalBlochEquations}
    \begin{align}
        \frac{d}{dt}\rho_{10} &= -i\frac{g}{2} (1-w)\rho_{10} + i \Omega w,\\
        \frac{d }{dt}w &= -4 \Omega {\mathrm{Im}} (\rho_{10}),
    \end{align}
\end{subequations}
where $g = n U (1-\kappa)/\hbar$. As the evolution is purely Hamiltonian, these equations conserve the expectation value of the Hamiltonian. In particular, as the number of atoms is also conserved, the mean energy per particle given by
\begin{align}
    E &= \frac{\mean{\hat{H}}}{\mean{\hat{N}}} = -\frac{1}{8}\hbar g(1 - w)^2 + 2 \hbar \Omega {\mathrm{Re}} (\rho_{10})
    \label{eqOpticalBlochEnergy}
\end{align}
is conserved. The solutions to Eq.~(\ref{eqOpticalBlochEquations}) are plotted in Figure \ref{figBlochSphere}.

As the evolution is purely Hamiltonian, the state can be described by a point on the surface of the Bloch sphere (see Figure \ref{figBlochSphere}).  The state is however not completely free to move on this sphere as the energy per particle $E$ must be conserved by its motion. Equation (\ref{eqOpticalBlochEnergy}) is a holonomic constraint, which together with the identity $w^2 + 4 |\rho_{10}|^2=1$ reduce the number of degrees of freedom of the solution from three $(w, 2\,{\mathrm{Re}}(\rho_{10}), 2 \,{\mathrm{Im}}(\rho_{10})$) to one, restricting the system's motion to a one-dimensional subset of the surface of the Bloch sphere
(lines of constant color in Figure \ref{figBlochSphere}).

There exists an extension to the Poincar{\'{e}}-Bendixson theorem which states that for certain manifolds, including the sphere $S^2$, all possible paths traced out by a $C^2$ action (which includes paths in phase space from a sufficiently smooth Hamiltonian) must approach or be periodic cycles, go to fixed points, or be a curve that covers the entire surface \cite{Schwartz1963}. On a sphere the last possibility is clearly impossible due to the hairy ball theorem demonstrating the existence of at least one fixed point \cite{Eisenberg1979}.

In Appendix \ref{appNoLimitCyles} we show that the fixed points in the current system are unreachable, meaning they either cannot be reached in normal evolution, or are a trivial case of periodicity where nothing changes. The final possibility consistent with the theorem proved in \cite{Schwartz1963} is that some trajectories may approach a limit cycle. In Appendix \ref{appNoLimitCyles} we also show that limit cycles cannot exist in this system. Consequently, all trajectories in this system must be periodic.

Although it follows from the periodicity of the evolution of the state on the Bloch sphere that all physical expectation values are periodic (as they must be independent of the global phase), the global phase rotation must be cancelled for the wave functions themselves to be periodic. This can be achieved by appropriate choice of the energy offset $\mu$. It is the periodicity of the order parameters $\Psi_i$ that will enable the stability of the condensate to excitations to be determined.

\section{Excitation dynamics}
\label{SectionExcitationDynamics}

The evolution of small perturbations about the mean field dynamics of a condensate define both the excitation spectrum of the condensate and its stability to perturbations. 

To determine the evolution of these excitations the mean field dynamics must be separated from that of the excitations. As in Section~\ref{secBogOverview} we proceed by defining the deviation operators $\delta\hat{\Psi}_i = \hat{\Psi}_i - \mean{\hat{\Psi}_i}$ and treat $\delta\hat{\Psi}_i$ as a small quantity. In this case, however, the $\mean{\hat{\Psi}_i}=\Psi_i$ are themselves time dependent, obeying Eqs.~(\ref{eqMeanFieldEquationsOfMotion}) for the mean field.

The equations of motion for the deviation operators are obtained by replacing the field operators $\hat{\Psi}_i$ in the operator evolution equations (\ref{eqOperatorEquationsOfMotion}) with $\Psi_i + \delta \hat{\Psi}_i$ and keeping only terms up to first order in the deviation operators. Applying this procedure gives
\begin{subequations}
    \label{eqDeviationOperatorsEvolutionXSpace}
    \begin{align}
        \begin{split}
            i \hbar \frac{\partial }{\partial t}\delta \hat{\Psi}_1 =& U \left[ \left(2 |\Psi_1|^2 + |\Psi_0|^2 \right)\delta \hat{\Psi}_1 + \Psi_1^2\delta\hat{\Psi}_1^\dagger \right. \\
&+ \left. \Psi_1\Psi_0\delta\hat{\Psi}_0^\dagger + \Psi_0^*\Psi_1\delta\hat{\Psi}_0  \right] \\
                    &  - \frac{\hbar^2}{2M}\nabla^2 \delta \hat{\Psi}_1 +\hbar \Omega\delta\hat{\Psi}_0- \mu \delta \hat{\Psi}_1,
        \end{split}\\
        \begin{split}
        i \hbar \frac{\partial }{\partial t}\delta \hat{\Psi}_0 = & U\left[\left(2\kappa |\Psi_0|^2 + |\Psi_1|^2\right)\delta\hat{\Psi}_0 + \kappa \Psi_0^2 \delta\hat{\Psi}_0^\dagger\right. \\
& \left.  + \Psi_1\Psi_0\delta\hat{\Psi}_1^\dagger + \Psi_1^*\Psi_0\delta\hat{\Psi}_1 \right] \\
                    & -\frac{\hbar^2}{2M}\nabla^2 \delta \hat{\Psi}_0 +\hbar \Omega\delta\hat{\Psi}_1 - \mu \delta\hat{\Psi}_0.
        \end{split}
    \end{align}
\end{subequations}

Having assumed the mean field (but not the fluctuations) to be homogeneous, the evolution equations are spatially translation-invariant and take their simplest form in a Fourier basis. Performing the Fourier transform of Eqs.~(\ref{eqDeviationOperatorsEvolutionXSpace}) yields
\begin{subequations}
\label{eqDeviationOperatorsEvolutionKSpace}
\begin{eqnarray}
i \hbar \frac{\partial }{\partial t}\delta \hat{\Psi}_1(\mathbf{k}) &=& U \Big[ \left(2 |\Psi_1|^2 + |\Psi_0|^2 \right)\delta \hat{\Psi}_1(\mathbf{k})  \nonumber \\
	&&  + \Psi_1^2\delta\hat{\Psi}_1^\dagger(-\mathbf{k})+ \Psi_1\Psi_0\delta\hat{\Psi}_0^\dagger(-\mathbf{k}) \nonumber \\
	&& + \Psi_0^*\Psi_1\delta\hat{\Psi}_0(\mathbf{k}) \Big] +\frac{\hbar^2 \mathbf{k}^2}{2M} \delta \hat{\Psi}_1(\mathbf{k})  \nonumber \\
	&& +\hbar \Omega\delta\hat{\Psi}_0(\mathbf{k})- \mu \delta \hat{\Psi}_1(\mathbf{k}), \\
i \hbar \frac{\partial}{\partial t} \delta \hat{\Psi}_0(\mathbf{k}) &=& U \Big[ \left(2\kappa |\Psi_0|^2 + |\Psi_1|^2\right) \delta\hat{\Psi}_0(\mathbf{k}) \nonumber \\
	&& + \kappa \Psi_0^2 \delta\hat{\Psi}_0^\dagger(-\mathbf{k}) + \Psi_1\Psi_0\delta\hat{\Psi}_1^\dagger(-\mathbf{k}) \nonumber \\
	&& + \Psi_1^*\Psi_0\delta\hat{\Psi}_1(\mathbf{k}) \Big] +\frac{\hbar^2 \mathbf{k}^2}{2M} \delta \hat{\Psi}_0(\mathbf{k}) \nonumber \\
	&&  +\hbar \Omega\delta\hat{\Psi}_1(\mathbf{k}) - \mu \delta\hat{\Psi}_0(\mathbf{k}).
\end{eqnarray}
\end{subequations}

In this form, it is clear that the Fourier modes are almost completely decoupled from each other. Each deviation operator $\delta\hat{\Psi}_i(\mathbf{k})$ is only coupled to $\left\{\delta\hat{\Psi}_j(\mathbf{k}),\, \delta\hat{\Psi}_j^\dagger(-\mathbf{k})\right\}$, with each $\delta\hat{\Psi}_i^\dagger(-\mathbf{k})$ also only coupled to this same set. This can be exploited to write Eqs.~(\ref{eqDeviationOperatorsEvolutionKSpace}) in matrix form as
\begin{align}
    \label{eqDeviationOperatorsMatrixEvolution}
    i \hbar \frac{\partial }{\partial t}\hat{\bm{\Upsilon}}(\mathbf{k}) &= \mathcal{H}(\mathbf{k}) \hat{\bm{\Upsilon}}(\mathbf{k}),
\end{align}
where
\begin{equation}
    \hat{\bm{\Upsilon}}(\mathbf{k}) = 
    \begin{pmatrix}
        \delta\hat{\Psi}_1(\mathbf{k}) &
        \delta\hat{\Psi}_1^\dagger(-\mathbf{k}) &
        \delta\hat{\Psi}_0(\mathbf{k}) &
        \delta\hat{\Psi}_0^\dagger(-\mathbf{k})
    \end{pmatrix}^\text{T},
\end{equation}
\begin{widetext}
\begin{equation}
    \mathcal{H}(\mathbf{k}) = 
    \begin{pmatrix}
        \varepsilon(\mathbf{k}) + q_{1} - \mu & v_{11} & u_{01} + \hbar \Omega & v_{10}\\
        -v_{11}^* & -\varepsilon(\mathbf{k}) - q_1 + \mu & -v_{10}^* & -u_{10} - \hbar \Omega\\
        u_{10} + \hbar \Omega & v_{10} & \varepsilon(\mathbf{k}) + q_0 - \mu & \kappa v_{00}\\
        -v_{10}^* & -u_{01} - \hbar \Omega & -\kappa v_{00}^* & -\varepsilon(\mathbf{k}) - q_0 + \mu
    \end{pmatrix},\label{eqHMatrix}
\end{equation}
\end{widetext}
and $q_1 = U\left(2 |\Psi_1|^2 + |\Psi_0|^2\right)$, $q_0 = U\left(2\kappa |\Psi_0|^2 + |\Psi_1|^2\right)$, $u_{ij} = U\Psi_i^*\Psi_j$, $v_{ij} = U\Psi_i\Psi_j$, and $\displaystyle \protect{\varepsilon(\mathbf{k}) = \hbar^2 \mathbf{k}^2 / 2M}$.

Note that the matrix $\mathcal{H}(\mathbf{k})$ is not the Hamiltonian, but is related to it by Eq.~(\ref{eqScriptHRelationshipToHamiltonian}). As a consequence, although it will be shown later that in some circumstances $\mathcal{H}(\mathbf{k})$ contains complex eigenvalues and is hence not Hermitian, this in no way conflicts with the requirement that the Hamiltonian $\hat{H}$ must be Hermitian and only have real eigenvalues.

If the coefficients of the matrix $\mathcal{H}(\mathbf{k})$ were not time-dependent, the excitation spectrum of the condensate could simply be obtained from the eigenvalues of $\mathcal{H}(\mathbf{k})$. Non-zero imaginary components for these eigenvalues would indicate the corresponding mode to be unstable. This is true independent of the sign of the imaginary component as the eigenvalues come in pairs with opposite imaginary components. A full discussion of this issue is given in Appendix \ref{appExcitationEvolution}.

Before considering the general case of $\kappa \neq 1$, it is instructive to consider the limit in which all scattering lengths are equal ($\kappa=1$) and recover some familiar results. If we assume $\kappa = 1$, the nonlinear term in Eq.~(\ref{eqMeanFieldEquationsOfMotion}) only contributes to a rotation of the global phase of the spinor condensate. In this case the dynamics can be solved analytically and familiar excitation spectra recovered. We find that the general solution is
\begin{subequations}
    \label{eqKappa1MeanFieldSolution}
    \begin{align}
        \Psi_1(t) &= \cos(\Omega t) \Phi_+ + \sin(\Omega t) \Phi_-, \\
        \Psi_0(t) &= -i\sin(\Omega t) \Phi_+ + i\cos(\Omega t) \Phi_-,
    \end{align}
\end{subequations}
for some complex constants $\Phi_\pm$, and where the chemical potential $\mu = n U$ has cancelled the global phase rotation. This solution can be viewed as a linear basis transformation from $\hat{\Psi}_i$ to $\hat{\Phi}_\pm$, which are the eigenvectors of the Rabi coupling term in Eq.~(\ref{eqInitialHamiltonian}). Performing this change of basis on the original Hamiltonian given by Eq.~(\ref{eqInitialHamiltonian}) yields a Hamiltonian of the same form, but without the Rabi coupling term. The equations of motion for the deviation operators $\delta\hat{\Phi}_\pm$ therefore give a matrix of precisely the same form as Eq.~(\ref{eqHMatrix}) but in terms of $\Phi_\pm$ and $\delta\hat{\Phi}_\pm$ instead of the $\Psi_i$ and $\delta\hat{\Psi}_i$, and with $\Omega$ replaced by 0. This new matrix $\mathcal{H}'(\mathbf{k})$ is time-independent and can be diagonalized to give the eigenvalues
\begin{subequations}
    \label{eqKappa1Eigenvalues}
    \begin{align}
        \hbar \omega_\uparrow(\mathbf{k}) &= \sqrt{\varepsilon(\mathbf{k})\left(\varepsilon(\mathbf{k}) + 2 n U\right)},\\
        \hbar \omega_\downarrow(\mathbf{k}) &= \varepsilon(\mathbf{k}),
    \end{align}
\end{subequations}
where $\varepsilon(\mathbf{k}) = \hbar^2\mathbf{k}^2 / 2M$, $n= |\Psi_1|^2 + |\Psi_0|^2$ and the remaining two eigenvalues are the negatives of Eq.~(\ref{eqKappa1Eigenvalues}). $\hbar \omega_\uparrow(\mathbf{k})$ is the usual Bogoliubov spectrum \cite{Bogoliubov1947} corresponding to excitations in the total condensate density. The eigenvalue $\hbar \omega_\downarrow(\mathbf{k})$ is the free particle spectrum; this excitation only changes the relative densities of the two states without affecting the total density, hence not affecting the nonlinear term in the Hamiltonian given by Eq.~(\ref{eqInitialHamiltonian}).

As explained in Section~\ref{secBogOverview}, the Hamiltonian for the condensate excitations that corresponds to the eigenvalues given by Eq.~(\ref{eqKappa1Eigenvalues}) is
\begin{align}
    \hat{H} &= \sum_{i=\uparrow,\downarrow}\int d\mathbf{k}\, \hbar \omega_i(\mathbf{k}) \hat{\Lambda}_i^\dagger(\mathbf{k}) \hat{\Lambda}_i^{\phantom{\dagger}}(\mathbf{k}),
\end{align}
where the $\hat{\Lambda}_{\uparrow,\downarrow}(\mathbf{k})$ obey boson commutation relations and are the corresponding normalised eigenvectors to the eigenvalues in Eq.~(\ref{eqKappa1Eigenvalues}). The normalised eigenvectors for the negatives of those eigenvalues are the $\hat{\Lambda}_{\uparrow, \downarrow}^\dagger(-\mathbf{k})$.

\section{Instabilities and the excitation spectrum}
\label{secInstabilitiesAndExcitations}

Having considered the limit of equal scattering lengths, it now remains to determine the energy spectrum and condensate stability in the general case of $\kappa \neq 1$. 


In the general case, the excitation spectrum cannot be obtained from the eigenvalues of the matrix $\mathcal{H}(\mathbf{k})$ in Eq.~(\ref{eqHMatrix}) as the matrix's entries are themselves time-dependent. However, due to the periodicity of the mean field wave functions demonstrated in Section~\ref{secPeriodicSolutions}, the entries of $\mathcal{H}(\mathbf{k})$ are themselves periodic, which enables Floquet's theorem to be applied.

Floquet's theorem proves that the matrix solution to the initial-value problem
\begin{subequations}
    \label{eqFloquetMatrixIVP}
    \begin{align}
        \frac{d}{dt}\bm{\Pi}(t) &= \bm{A}(t) \bm{\Pi}(t),\\
        \bm{\Pi}(0) &= \mathbb{I},
    \end{align}
\end{subequations}
where $\mathbb{I}$ is the $n \times n$ identity matrix and $\bm{A}(t)$ a periodic $n \times n$ matrix with period $T$ can be written in the form
\begin{align}
    \bm{\Pi}(t) = \bm{P}(t) \exp(-i\bm{Q} t),
    \label{eqFloquetSolution}
\end{align}
where $\bm{Q}$ is some constant matrix, $\bm{P}(t)$ is a matrix of periodic functions with period $T$, and $\bm{P}(0) = \mathbb{I}$.

We note there are differing definitions of the matrix $\bm{Q}$. While it is usual in quantum mechanics literature \cite{Shirley1965,Hanggi1998,Garrison1999} to define Eq.~(\ref{eqFloquetSolution}) with the $-i$ in the exponent, in mathematics literature the $-i$ is omitted (see, for example \cite{Moulton1958,AppliedNonlinearDynamics}).

The matrix solution $\bm{\Pi}(t)$ is the general solution to the related linear system
\begin{align}
    \frac{d}{dt}\bm{x}(t) &= \bm{A}(t) \bm{x}(t),
    \label{eqFloquetVectorIVP}
\end{align}
for any initial condition $\bm{x}(0)$ where $\bm{x}(t)$ is a vector. Every solution $\bm{x}(t)$ to this problem can be written in terms of the matrix $\bm{\Pi}(t)$ using
\begin{align}
    \bm{x}(t) &= \bm{\Pi}(t) \bm{x}(0),
\end{align}
as is easily verified. The matrix solution $\bm{\Pi}(t)$ thus completely determines the behaviour of all solutions to Eq.~(\ref{eqFloquetVectorIVP}), or, equivalently, Eq.~(\ref{eqDeviationOperatorsMatrixEvolution}).

The eigenvalues of $\bm{Q}$ are known as \emph{Floquet exponents} (or \emph{characteristic exponents}) and determine the long-term growth or decay of the solutions to Eq.~(\ref{eqFloquetVectorIVP}). These eigenvalues can be obtained from the \emph{monodromy matrix},
\begin{align}
    \label{eqMonodromyMatrix}
    \mathcal{M} &= \bm{\Pi}(T) = \exp(-i\bm{Q} T),
\end{align}
as $\bm{P}(T) = \bm{P}(0) = \mathbb{I}$. The existence and uniqueness of the solution to Eq.~(\ref{eqFloquetMatrixIVP}) guarantees that $\bm{\Pi}(t)$ and hence $\mathcal{M}$ will be invertible. The Floquet exponents $\xi_i$ can therefore be obtained from the eigenvalues $\lambda_i$ of the monodromy matrix using $\lambda_i = \exp(-i\xi_i T)$.
It is the Floquet exponents of the matrix $\mathcal{H}(\mathbf{k})$ that we wish to calculate in order to determine the stability of the condensate to excitations, as these exponents will determine whether the long term behaviour of the system exhibits exponential growth or decay.

Determining the Floquet exponents of the system via Eq.~(\ref{eqDeviationOperatorsMatrixEvolution}) requires knowledge of its period $T$, and hence the period of the mean field dynamics given by Eq.~(\ref{eqMeanFieldEquationsOfMotion}). Although this period cannot generally be determined analytically, it can be found numerically.

It was shown in Section~\ref{secPeriodicSolutions} that up to a global phase rotation $\displaystyle f(T) = e^{i 2\pi \Delta \nu T}f(0)$, the mean fields $\Psi_j(t)$ are periodic. The mean fields $\Psi_j(t)$ can be therefore written in the form
\begin{align}
    \label{eqMeanFieldFourierDecomposition}
    \Psi_j(t) = \sum_{n=-\infty}^\infty \alpha_{j,n} \exp\left[i 2\pi \left( n \nu_0 + \Delta\nu\right)t \right],
\end{align}
for some complex constants $\alpha_{j, n}$, fundamental frequency $\nu_0 = T^{-1}$, and frequency offset $\Delta \nu$. In this form, the $\Psi_j(t)$ are not exactly periodic as $\Psi_j(T) = \exp(i\Delta \nu T)\Psi_j(0)$, but this frequency offset can be cancelled by an appropriate choice of the energy offset $\mu = -2\pi\hbar \Delta \nu$ in Eq.~(\ref{eqInitialHamiltonian}).

The period $T$ and frequency offset $\Delta\nu$ in Eq.~(\ref{eqMeanFieldFourierDecomposition}) can be determined from the Fourier transform of $\Psi_j(t)$, which will have sharp peaks at the frequencies $n \nu_0 + \Delta \nu$. Choosing the energy offset $\mu= - 2 \pi \hbar \Delta\nu$, the frequency offset in Eq.~(\ref{eqMeanFieldFourierDecomposition}) can be cancelled making the $\Psi_j(t)$ with this energy offset exactly periodic with period $T$.

As an example of this process, we consider a two-level condensate with physical parameters corresponding to those used in the metastable helium BEC experiment by Dall \emph{et al.}\ \cite{Dall2009}. While a more detailed analysis of this experiment using our theory will be done in Section~\ref{secTheoryExperimentAgreement}, we note that it fulfils the criteria we require --- a multi-state, high aspect ratio condensate with the states possessing dissimilar scattering lengths and coupled by radio frequency radiation. Specifically we take a particle number of $N=2\times 10^6$, radial and axial trapping frequencies of $\omega_r = 2\pi \times 1020$Hz and $\omega_z = 2\pi \times 55$Hz respectively, choose a coupling field with a Rabi frequency of $\Omega = 2\pi\times 3$kHz, and nonlinearity mismatch parameter of $\kappa = 0.74$.

\begin{figure}
    \centering
    \includegraphics[width=9cm]{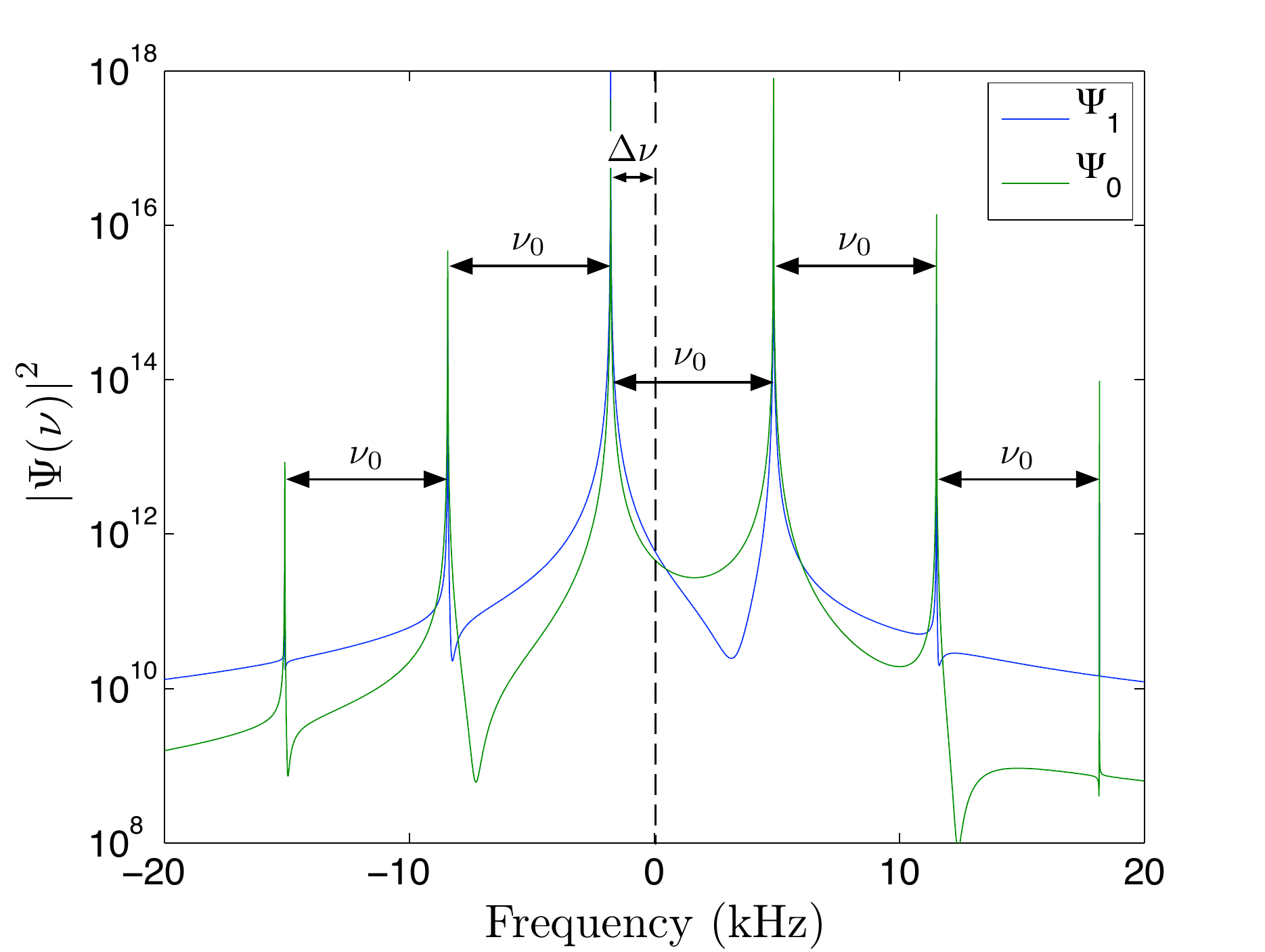}
    \caption{(color online) Temporal Fourier transform of the calculated mean field evolution defined by Eq.~(\ref{eqMeanFieldEquationsOfMotion}) for the centre of a two-state metastable helium condensate $N=2\times 10^6$, $\omega_r = 2\pi \times 1020$Hz, $\omega_z = 2\pi \times 55$Hz, $\Omega = 2\pi\times 3$kHz, $\kappa = 0.74$. The frequency $\nu_0$ is the inverse period of the system, and $\Delta\nu$ represents the global phase rotation. From the data in this figure the values $\nu_0 = 6.65 $kHz and $\Delta\nu = -1.78$ kHz can be determined, giving the period as $T=150 \mu$s.} 
\label{figMeanFieldFourierTransform}
\end{figure}

We simulate the evolution of the mean field of such a BEC using Eqs.~(\ref{eqMeanFieldEquationsOfMotion}), and obtain a time series of the two condensate fields $\Psi_j(t)$ at the center of the condensate. We then Fourier transform these time series and obtain a power spectrum for the two fields, which is plotted in Figure~\ref{figMeanFieldFourierTransform}. The periodic behaviour of the fields is clearly visible, and we can easily extract the period of $T=150\mu$s and the frequency offset of $\Delta \nu = -1.78$kHz.

\begin{figure}
    \centering
    \includegraphics[height=14cm]{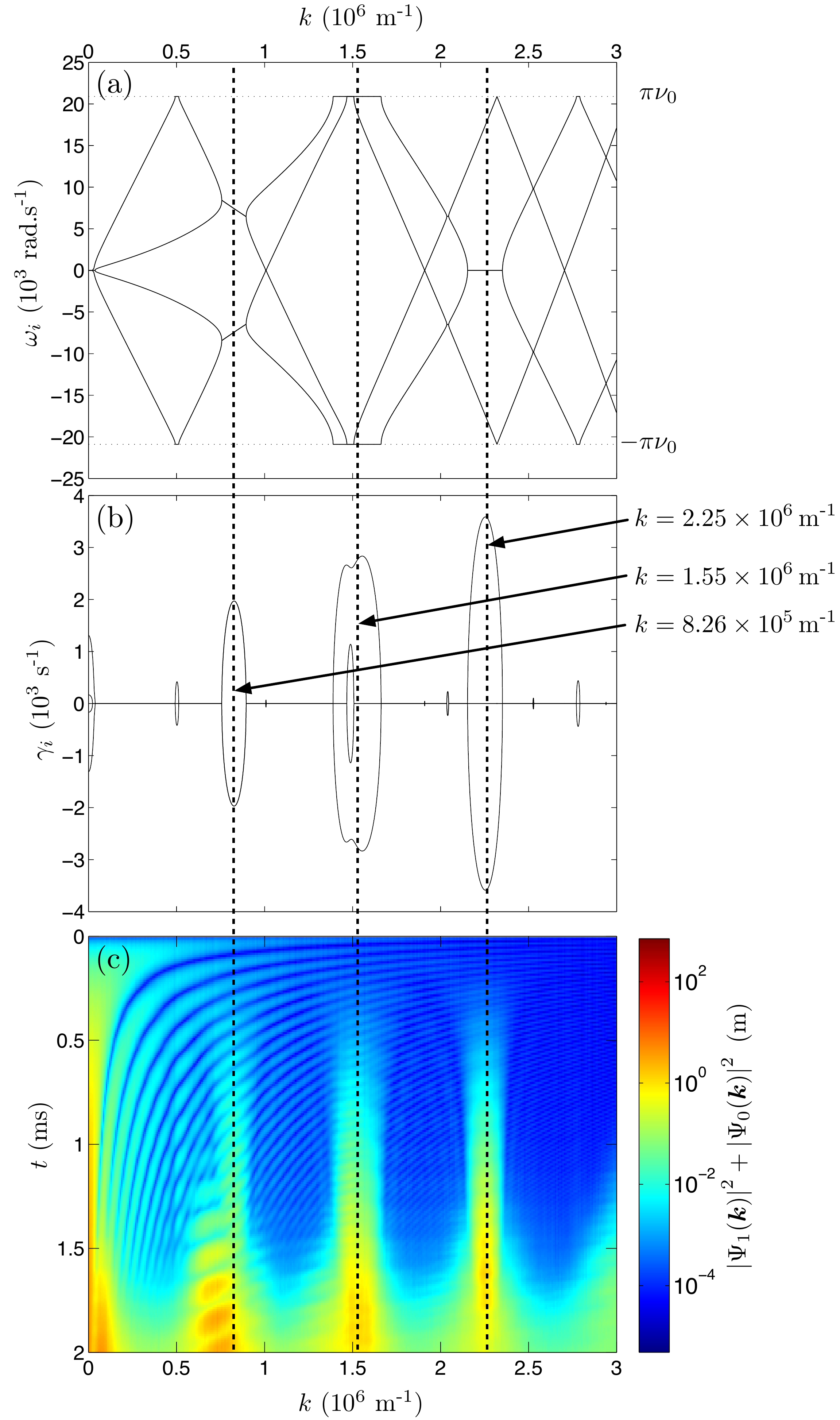}
    \caption{(color online) Illustration of the Floquet exponents for $\mathcal{H}(\mathbf{k})$ and comparison to a corresponding truncated Wigner simulation for $\Omega = 2\pi \times 3$kHz.
        Upper figure (a) displays the real part $\omega_i$ of the Floquet exponents. As the $\omega_i$ can only be determined up to a multiple of $2\pi\nu_0$ (see Eq.~(\ref{eqAmbiguityFloquetExponent})), they have been reduced modulo $2\pi\nu_0$ into the range $\left[-\pi\nu_0,\, \pi\nu_0\right]$.
        The middle figure (b) shows the imaginary part $\gamma_i$ of the Floquet exponents, which indicate an instability for the corresponding wave number when they are non-zero. Note that what appears to be a horizontal line at $\gamma_i=0$ is not an axis, but a plot of the $\gamma_i(k)$, which are mostly zero. Thus at $k=8.26\times 10^{5}$m$^{-1}$ all four modes are unstable, but at $k=2.25\times 10^{6}$m$^{-1}$ only two modes are unstable. As discussed in the Section~\ref{secDiscussion}, the two-mode regime is more suitable for experimental detection of entanglement due to the entanglement being between fewer modes.
        Lower figure (c) shows the results of a 1D truncated Wigner simulation corresponding to the system under consideration in (a) and (b). The truncated Wigner simulation exhibits growth in the same modes predicted from the results of the perturbative analysis shown in (b). The truncated Wigner results shown are the average of 500 realisations.
        \label{figCondensateEigenvalues}}
\end{figure}

The period and energy offset determined, it remains to calculate the monodromy matrix $\mathcal{M}(\mathbf{k})$ from which the Floquet exponents may be derived. This is achieved by numerically solving the related matrix problem to Eq.~(\ref{eqDeviationOperatorsMatrixEvolution}) from $t=0$ to $t=T$, as can be seen from Eq.~(\ref{eqMonodromyMatrix}). Noting that the matrix $\mathcal{H}(\mathbf{k})$ only depends on $k = |\mathbf{k}|$, the solutions for the Floquet exponents $\xi(k) = \omega(k) + i\gamma(k)$ are shown in Figure~\ref{figCondensateEigenvalues}.

In the limit that the mean field is time-independent (a degenerate case of periodicity), the Floquet exponents $\xi(\mathbf{k})$ are related to the eigenvalues $\lambda(\mathbf{k})$ of the matrix $\mathcal{H}$ by $\xi(\mathbf{k}) = \lambda(\mathbf{k})/\hbar$. It was previously stated that eigenvalues of $\mathcal{H}(\mathbf{k})$ with non-zero imaginary components would be unstable, this is also true for the Floquet exponents $\xi(\mathbf{k})$.

The temporal periodicity of the system implies that the real components $\omega_i(\mathbf{k})$ of the Floquet exponents are only uniquely defined modulo $2\pi \nu_0$. This is because any eigenvalue $\lambda$ of the monodromy matrix $\mathcal{M}(\mathbf{k})$ corresponds to infinitely many Floquet exponents,
\begin{align}
    \label{eqAmbiguityFloquetExponent}
    \lambda = \exp\left[-i \left(\omega + 2 \pi n \nu_0 + i \gamma\right)T\right] = \exp\left[-i\left(\omega + i \gamma\right)T\right].
\end{align}
This does not hinder our understanding of the system as our primary interest is in the stability of the condensate to excitations, which is determined by the $\gamma_i(\mathbf{k})$.

The normalised eigenvectors of the system are not always annihilation or creation operators; as is discussed in Appendix \ref{appExcitationEvolution}, this is only true when the Floquet exponents are purely real. When the Floquet exponents have a non-zero imaginary component, annihilation and creation operators can be constructed from linear combinations of the eigenvectors. Not being eigenvectors, these operators will therefore have non-trivial evolution. As is shown in Appendix \ref{appExcitationEvolution}, Floquet exponents with non-zero imaginary parts come in pairs of the form $ \omega(\mathbf{k}) \pm i \gamma(\mathbf{k})$. From the corresponding eigenvectors to these Floquet exponents the bosonic annihilation operator $\hat{\Lambda}(\mathbf{k}, t)$ can be formed, which evolves as
\begin{subequations}
    \label{eqDynamicalInstability}
    \begin{align}
        \hat{\Lambda}(\mathbf{k}, nT) = e^{i n\omega(\mathbf{k}) T} &\left( \sinh(n\gamma(\mathbf{k}) T) \hat{\Lambda}'^\dagger(-\mathbf{k}, 0) \right. \nonumber \\
    & \left. + \cosh(n\gamma(\mathbf{k}) T) \hat{\Lambda}(\mathbf{k}, 0)\right),\\
        \hat{\Lambda}'(\mathbf{k}, nT) = e^{-i n \omega(\mathbf{k}) T} &\left( \sinh(n\gamma(\mathbf{k}) T) \hat{\Lambda}^\dagger(-\mathbf{k}, 0)  \right. \nonumber \\
   &+ \left. \cosh(n\gamma(\mathbf{k}) T) \hat{\Lambda}'(\mathbf{k}, 0)\right),
    \end{align}
\end{subequations}
where $n$ is a positive integer, and where $\hat{\Lambda}'(\mathbf{k}, t)$ will be equal to $\hat{\Lambda}(\mathbf{k}, t)$ in some circumstances as discussed in Appendix \ref{appExcitationEvolution}. Due to the exponential growth in Eq.~(\ref{eqDynamicalInstability}), the mode corresponding to $\hat{\Lambda}(\mathbf{k})$ represents a dynamical instability of the condensate.

The behaviour of the dynamical instabilities is governed by Eq.~(\ref{eqDynamicalInstability}) only while the unstable modes have a small occupation compared to the condensate, and scattering between the unstable modes can be neglected. Figure~\ref{figCondensateEigenvalues}(c) shows the results of a truncated Wigner simulation of the Hamiltonian Eq.~(\ref{eqInitialHamiltonian}), which is in excellent agreement with the location of the dynamical instabilities as determined by the Floquet exponents [see Figure~\ref{figCondensateEigenvalues}(b)]. For later times, there is an additional mode undergoing growth, $k \approx 7.5\times 10^5$m$^{-1}$. This mode is the result of scattering between the dynamical instabilities, a process neglected the perturbative approach taken in this section.  The depletion of the condensate mode has also been neglected in this analysis.  

In summary, the procedure used to find the Floquet exponents of $\mathcal{H}(\mathbf{k})$, and hence the stability of the condensate to excitations is:
\begin{enumerate}
    \item Numerically solve Eq.~(\ref{eqMeanFieldEquationsOfMotion}) with $\mu=0$ for a long time $t \gg T$ where $T$ is the periodicity of the solution.
    \item Perform the temporal Fourier transform of the solutions obtained for $\Psi_i(t)$ to accurately determine the period $T$ and the value of the energy offset $\mu$ required to cancel any global phase evolution to make the wave functions themselves periodic.
    \item Using the calculated period and energy offset, numerically solve the related matrix problem given by Eq.~(\ref{eqDeviationOperatorsMatrixEvolution}) for a range of values of $\mathbf{k}$ to obtain the monodromy matrix $\mathcal{M}(\mathbf{k})$. In this calculation the matrix $\mathbf{A}(t)$ in Eq.~(\ref{eqFloquetMatrixIVP}) is $-i\mathcal{H}(\mathbf{k}, t)/\hbar$.
    \item Calculate the eigenvalues $\lambda_i$ of $\mathcal{M}(\mathbf{k})$ and determine the Floquet exponents $\xi_i$ using $\protect{\lambda_i = \exp(-i \xi_i T)}$. The real parts of these Floquet exponents give the energy spectrum, with non-zero imaginary parts giving the growth rate for the corresponding instability.
\end{enumerate}

\section{Agreement of theory with experiment}
\label{secTheoryExperimentAgreement}

We have shown that if we apply a coupling between the two levels of a two-state condensate, then, provided the nonlinearity mismatch parameter $\kappa$ given by Eq.~(\ref{eqKappaDefinition}) is not equal to one, there exist instabilities which can result in excitations that grow exponentially at well-defined momentum values $\hbar \mathbf{k}$. These excitations, however, will only become amplified provided their de Broglie wavelength is smaller than the extent of the condensate in the direction of $\mathbf{k}$ --- if the entire condensate covers only a fraction of a wavelength, clearly no amplification can take place.

To demonstrate the validity of our theory, we apply it to the experiment of Dall \emph{et al.} \cite{Dall2009}. In this experiment metastable helium atoms in the $m_F = +1$ Zeeman state were held in magnetic trap and evaporatively cooled to form a BEC. An RF coupling field was then applied, flipping the atoms from the $m_F=+1$ magnetically trapped state to the $m_F=0$ untrapped state, forming an atom laser. These untrapped atoms fell from the trap onto a position sensitive detector 4cm below and the resulting two-dimensional beam cross section was imaged. In addition to the expected atom laser profile, they also observed two anomalous beams of atoms that had been ejected in the weak trapping direction of the condensate. 

One complication is that metastable helium in the $F=1$ manifold possesses three possible Zeeman states, $m_F = +1, 0, -1$, and our theory was developed for two-state systems. As the $m_F = -1$ state is antitrapped, any atoms in this state leave the condensate very rapidly due to the combined effects of both the mean field repulsion and the magnetic field gradient. For the experiment in question, a classical particle in the centre of the condensate under the influence of the same effective potential experienced by the $m_F=-1$ atoms would reach a momentum equal to the momentum width of the condensate (and hence no longer be able to couple to the stationary atoms in the middle of the condensate) in $\sim 80\mu$s, significantly shorter than the inverse Rabi frequency of $\sim 300 \mu$s. This indicates that the population of the $m_F = -1$ state will be low, and have minimal influence on the dynamics of the system, and consequently we make an approximation where we ignore the $m_F = -1$ state.

The relevant experimental parameters were a particle number of $N=2\times 10^6$, and radial and axial trapping frequencies of $\omega_r = 2\pi \times 1020$Hz and $\omega_z = 2\pi \times 55$Hz respectively, and the RF outcoupling field corresponding to maximum anomalous particle production had a Rabi frequency of $\Omega = 2\pi\times 3$kHz. Metastable helium has s-wave scattering lengths of $a_{1,1} = a_{1,0} = 7.51$nm and $a_{0,0}=5.56$nm \cite{Leo2001}, where $a_{i,j}$ is the scattering length between the two Zeeman states $m_F=i,j$, giving a nonlinear mismatch parameter of $\kappa = 0.74$.

We took two approaches to numerically modelling this system. The first was to simply solve the coupled Gross-Pitaevskii (GP) equations, which gives the mean field behaviour of the BEC and outcoupled atoms. The GP-equations, however, are a mean field approximation, and our theory involves physics beyond the mean field. Consequently the second approach was to numerically simulate the system using a phase-space method, specifically the truncated Wigner approximation (TWA). 

In both cases we simulated the full system in three dimensions in a fully multi-modal way, thus including spatial density variations across the condensate. This contrasts with the simpler numerical calculation carried out in Section~\ref{secInstabilitiesAndExcitations}, where we assumed a density that was constant in space but not time. As any real condensate has spatial density dependence, this is an important test of the validity of our model.

\begin{figure}
    \centering
    \includegraphics[width=9cm]{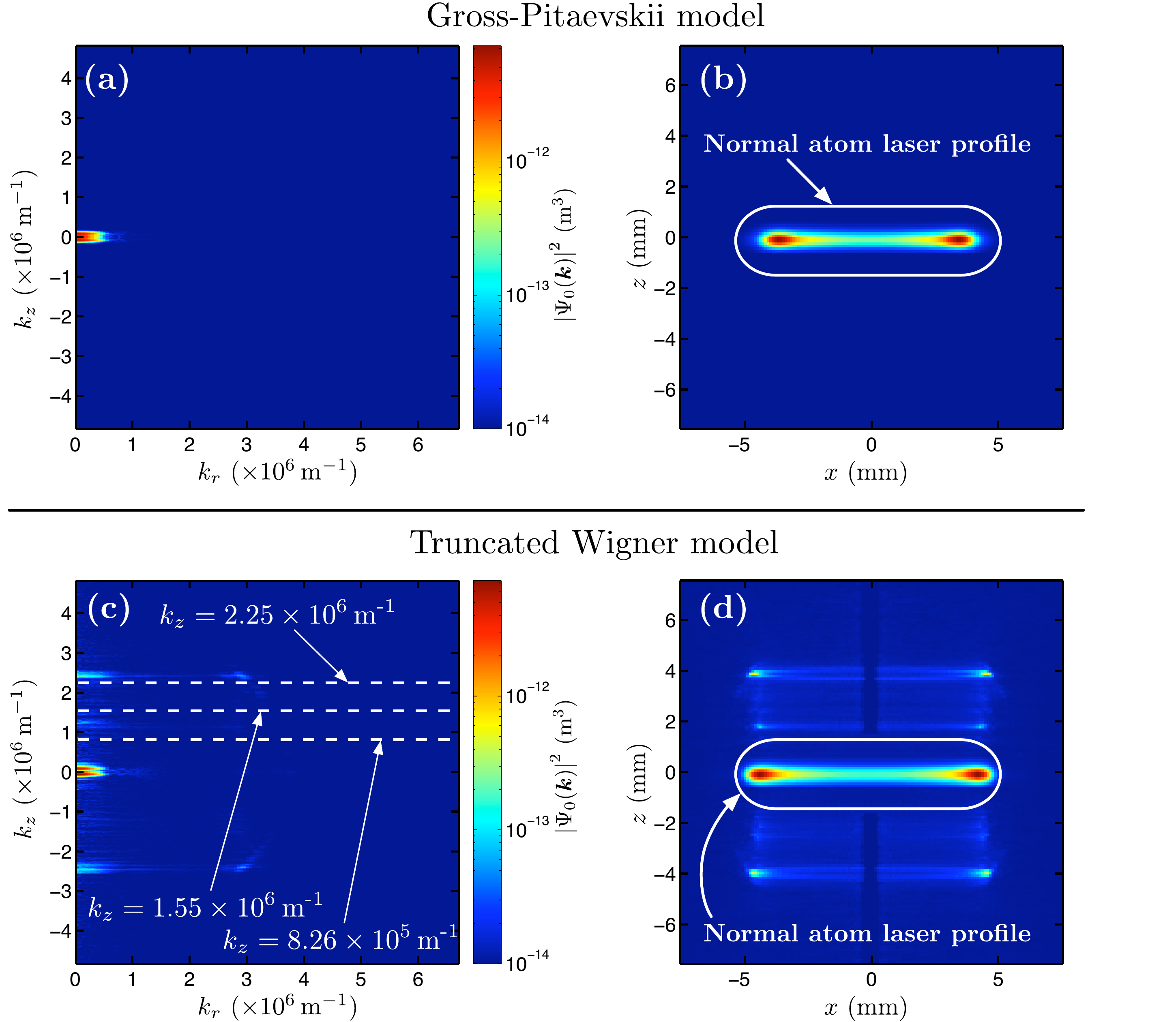}
    \caption{(color online) Simulation results for outcoupling from the centre of the condensate. Left figures (a) and (c) plot the momentum density of the untrapped state at $t=2$ ms as a function of the radial ($k_r$) and axial ($k_z$) wave numbers. Right figures (b) and (d) plot the normalised density profiles that would be observed on the MCP detector 4 cm below the condensate due to outcoupling from the condensate for $t=6$ ms. The upper figures (a) and (b) plot the results of a GP model where the $m_F=-1$ state is assumed negligible, lower figures (e) and (f) correspond to a Truncated Wigner simulation for the same system averaged over $N_\text{realisations} = 4$. The wave numbers of the three fastest-growing instabilities predicted by the perturbation analysis performed in Section~\ref{secInstabilitiesAndExcitations} are marked in (c). The weak axis is in the vertical direction in all figures.
\label{figTheoryZeroDetuningNoPIResults}}
\end{figure}

The results of these simulations are plotted in Figure~\ref{figTheoryZeroDetuningNoPIResults}, showing both the momentum distribution of the outcoupled atoms as well as the cross-sectional density profile of the atom laser on the detector. It is clear that the GP simulation shows no significant momentum components in the axial (weakly trapping, long condensate axis) direction. The TWA simulation on the other hand, shows significant specific momentum components have been excited along the axial direction. 

We have also plotted where we expect these momentum components to lie based on the perturbation analysis in Section~\ref{secInstabilitiesAndExcitations}. The dynamical instability with the largest growth rate illustrated in Figure~\ref{figCondensateEigenvalues} at $k=2.25\times 10^6$m$^{-1}$ is in good agreement with the highest growth rate instability in Figure~\ref{figTheoryZeroDetuningNoPIResults}(c) observed at $(k_r \approx 0,\, k_z \approx \pm 2.4\times 10^6$m$^{-1})$, and the position at which the particles corresponding to this instability land on the detector is in good agreement with the images from the experiment \cite{Dall2009}.

The absence of the effect in the GP simulations and the presence of the effect in the TWA simulations indicates that dynamic instabilities causing amplified excitations are indeed present in such systems, and can be experimentally created and measured.


\section{Existence of entanglement}
\label{secEntanglement}

While the production of oppositely directed matter wave beams though dynamic instabilities in a condensate is interesting, it is also worth noting that these beams should exhibit EPR entanglement on formation.

EPR entanglement was initially proposed from a first-quantized point of view \cite{Einstein1935}. Two individual particles are created with entanglement between two conjugate observables --- position and momentum, for example, or two orthogonal spins --- and then the particles are separated. Due to the entanglement, the measurement of one observable on one particle forces the other to instantaneously adopt a predictable value for the same observable, no matter how large the distance between the two particles.

In modern implementations, rather than considering discrete particles, one usually considers continuous variable entanglement between the quadratures of atomic or optical fields instead. Provided the correlation functions between the quadratures violate a specific inequality, the fields are EPR-entangled \cite{Reid1989}.

The standard procedure to create such fields in the optical regime is to use parametric down conversion \cite{WallsMilburn}. Here a single photon of frequency $2\omega$ interacts with a nonlinear crystal and is split into two photons of frequency $\omega$ in the degenerate case and two photons with frequencies $\omega_1 + \omega_2 = 2\omega$ in the nondegenerate case. The two generated photons are EPR correlated, and it is this process that has been the workhorse of almost all optical EPR tests. 

As discussed in Section~\ref{secInstabilitiesAndExcitations}, our scheme generates quasiparticles at specific momenta, and it is possible to derive the equations of motion for the bosonic operators describing these quasiparticles. Restated, these equations of motion are
\begin{subequations}
    \label{eqEOMforLambdaRestated}
    \begin{align}
        \hat{\Lambda}(\mathbf{k}, nT) = e^{i n\omega(\mathbf{k}) T} &\left( \sinh(n\gamma(\mathbf{k}) T) \hat{\Lambda}'^\dagger(-\mathbf{k}, 0) \right. \nonumber \\
    & \left. + \cosh(n\gamma(\mathbf{k}) T) \hat{\Lambda}(\mathbf{k}, 0)\right),\\
        \hat{\Lambda}'(\mathbf{k}, nT) = e^{-i n \omega(\mathbf{k}) T} &\left( \sinh(n\gamma(\mathbf{k}) T) \hat{\Lambda}^\dagger(-\mathbf{k}, 0)  \right. \nonumber \\
   &+ \left. \cosh(n\gamma(\mathbf{k}) T) \hat{\Lambda}'(\mathbf{k}, 0)\right).
    \end{align}
\end{subequations}
These equations describe the exponential amplification of two distinct modes, with each mode existing as a superposition of the $|1\rangle$ and $|0\rangle$ states. Equations~(\ref{eqEOMforLambdaRestated}) are identical to those describing optical two-mode parametric down-conversion \cite{WallsMilburn}, with the Floquet exponent $\gamma$ playing the role of the strength of the crystal nonlinearity. Just as these equations lead to EPR entanglement for photons, here they lead to EPR entanglement between the quasiparticle excitations. We note that Equations~(\ref{eqEOMforLambdaRestated}) do not strictly give the continuous time evolution of the atomic fields, but rather are only stated for times that are multiples of the Floquet period $T$. However as the Floquet exponents govern the long term behaviour of the system, the entanglement of the atomic system will asympotically approach that of the analogous optical system, even for times $t \neq T$.

The entanglement produced by the evolution described by Eq.~(\ref{eqEOMforLambdaRestated}) is between the oppositely-propagating modes $\hat{\Lambda}(\mathbf{k})$ and $\hat{\Lambda}'(-\mathbf{k})$.  Strictly, Eq.~(\ref{eqEOMforLambdaRestated}) only applies when all the real parts of all four Floquet exponents are of equal magnitude, and the imaginary components are of equal magnitude as is the case near $k = 8.26 \times 10^5\, \text{m\textsuperscript{-1}}$ in Figure~\ref{figCondensateEigenvalues}.  When only two Floquet exponents have non-zero imaginary components (or the real parts are of different magnitude), $\hat{\Lambda}'(\mathbf{k}) = \hat{\Lambda}(\mathbf{k})$, and the entanglement will be between $\hat{\Lambda}(\mathbf{k})$ and $\hat{\Lambda}(-\mathbf{k})$.  Equations~(\ref{eqEOMforLambdaRestated}) are consistent in this case as either $\omega(\mathbf{k}) = 0$ or $\omega(\mathbf{k}) = \pi \nu_0$, giving $\exp\left(i n \omega(\mathbf{k})T\right) = \pm 1$.

As the $\hat{\Lambda}$ and $\hat{\Lambda}'$ are superpositions of the $|1\rangle$ and $|0\rangle$ states, the entanglement will only be measured in the appropriate superpositions of these states.  For the case in which only two Floquet exponents have non-zero imaginary components (i.e.\ $\hat{\Lambda}(\mathbf{k}) = \hat{\Lambda}'(\mathbf{k})$) there will only be one superposition of the $|1\rangle$ and $|0\rangle$ states that will be formed at that momentum (the other being stable will not undergo exponential growth), and it will be the oppositely-propagating mode of this superposition with which it will be entangled.  In the case in which all four Floquet exponents have equal (non-zero) magnitude imaginary components and equal magnitude real components, the entanglement will be between \emph{orthogonal}, oppositely-propagating superpositions of the $|1\rangle$ and $|0\rangle$ states.  For example, if $\hat{\Lambda}(\mathbf{k}, 0)$ were purely the $|1\rangle$ state, it would be entangled with the $|0\rangle$ state propagating in the opposite direction.

It should also be noted that this analysis holds true both for the case where the $\hat{\Lambda}$ and $\hat{\Lambda}'$ are spontaneously seeded, and for the more general case in which the $\hat{\Lambda}$ and $\hat{\Lambda}'$ are seeded by different coherent states.

\section{Discussion and experimental practicality}
\label{secDiscussion}

We have demonstrated that our scheme can create particle beams along the long axis of a BEC, and shown that the generation of these these beams has the same mathematical form as optical non-degenerate parametric down conversion, resulting in EPR-entangled pairs of excitations with oppositely-directed momenta. In addition, an experiment satisfying the criteria of our scheme has been shown to exhibit cones of particles ejected along the long axis of a BEC, with momenta generally agreeing with our predictions.


It is clear that since the entangled modes are excitations described by the $\hat{\Lambda}(\mathbf{k})$ operators, it is not a simple case of, say, an $m_F=0$ atom being entangled with an $m_F=1$ atom with opposite momentum, as the entangled $\hat{\Lambda}(\mathbf{k})$ modes are each superpositions of $m_F=1$ and $m_F=0$ states. This is a problem in an experiment where the two states are not confined in the same potential, as in the metastable helium experiment discussed earlier. In such a case, the $m_F=1$ atoms remain magnetically confined, while the $m_F=0$ atoms can leave the condensate. However, if the $m_F=1$ components of the entangled modes are outcoupled faster than the time to undergo half an oscillation in the trap and reverse their momenta ($\sim 9\,$ms in the case of the helium experiment), then it should be possible to at least observe number difference squeezing between the atoms emitted in opposite directions. Even if the $m_F=1$ atoms have enough momentum to escape, however, the fact that they see a different potential to the $m_F=0$ atoms leads to considerable difference in their dynamics. 

One solution would be to use optically trapped atoms, with the Zeeman substates split via an external magnetic field. This will ensure that each substate will see the same confining potential. The difficulty with this approach is most of the conveniently accessible systems have at least three Zeeman states, ranging over $\{ -j, -j+1, \ldots j \}$. If RF radiation is used to couple these states, then since the energy splitting between the states is (usually) linear, all three states will become populated, breaking our two-level assumption.  One might think that in such a three state system one could couple $m_F=-1$ and $m_F=+1$ via a Raman transition, but such two-photon $\Delta m_F =2$ transitions within the same $F$ manifold are heavily suppressed \cite{hagleyET1999}. A better approach would be to use magnetically trapped BECs in two separate hyperfine manifolds, and couple two specific Zeeman states --- $|F=1,m_F =-1 \rangle$ and $|F=2,m_F =+1 \rangle$, for example.  Another approach could be to use the quadratic Zeeman shift to remove the degeneracy of the energy splittings in an $F$ manifold.  Using a sufficiently large bias field in conjunction with an optical trap would make RF coupling between the $m=1$ and $m=0$ levels of an $F=1$ manifold effectively a two-level system.

If a two-level system cannot be obtained, a three-level system could still be used. This is not a serious problem provided all atoms begin in a single trapped state (whichever one of the $m_F=\pm 1$ states is trapped in the magnetically trapped case, or either of these states in the optical trapping case), couple to the $m_F=0$ state, and ensure the coupling is weak enough that the atoms escape more quickly than the period associated with the Rabi frequency of the coupling. This will ensure that almost all of the population remains in two levels, and the system behaves like a two-level system. As discussed in Section~\ref{secTheoryExperimentAgreement}, this approximation describes the case for a magnetically trapped metastable helium BEC quite well, despite its three levels.

Although the particles forming the entangled beam will usually have enough energy to escape the trapping potential, it is possible that for some choice of atomic species, atomic density, and trapping geometry the momentum at which the effect occurs may not be be high enough. In this case the experiment could be run in pulsed mode, dropping the trapping potential after the entangled excitations have formed, and allowing them to escape. As these entangled particles leave along the weakly trapping axis, they will be well-resolved from the bulk of the condensate which expands most rapidly along the tight trapping axis.

The actual measurement of EPR quadrature entanglement requires the existence of a local oscillator to turn the quadrature entanglement into number correlations. Local oscillators are difficult to obtain for atomic fields; generally the only option is to use the source condensate itself. In this case the condensate would be split in two, with one of the condensates used to generate the entangled beams and the other used to generate the atomic beams that will form the local oscillator. As the condensates will initially only be displaced by a few hundred microns, the beams will overlap well in the far field allowing reasonable mode matching.

The most experimentally promising regime in which to observe entanglement would be to consider atomic beams for which only two Floquet exponents have non-zero imaginary components (this is observed, for example, near $k = 2.25\times 10^6\,\text{m\textsuperscript{-1}}$ in the He* experiment).  In this regime the two counter-propagating local oscillators will have the same superposition of $|1\rangle$ and $|0\rangle$ states, presenting less of a challenge than producing oppositely-propagating atomic beams in orthogonal superpositions.  Additionally the correct superposition of $|1\rangle$ and $|0\rangle$ states can be determined experimentally up to a relative phase difference by measuring the fraction of $|1\rangle$ and $|0\rangle$ atoms in the produced atomic beams.  If the correct superposition were not known in advance it would otherwise be necessary to scan over all of them, which is a two-dimensional space.  Knowing the ratio of $|1\rangle$ and $|0\rangle$ atoms would reduce the search space to a single dimension.

The regime in which all four Floquet exponents have non-zero imaginary components of equal magnitude (and equal magnitude real components) is interesting as two pairs of entangled counter-propagating beams are produced.  Measuring this entanglement is however a greater challenge as the local oscillators are more complicated, and it would require searching over a larger space to find the entanglement unless highly accurate theoretical predictions for the experiment were available.

The generation of the beams also relies on the correct choice of atomic species. The effect depends on a nonlinearity mismatch, requiring that the s-wave scattering lengths in the system are not all identical, so that $\kappa \neq 1$. This is not necessarily easy to achieve. $^{87}$Rb, for example, is a popular and well-understood atom in the BEC community, but if the $F=1$ hyperfine manifold is used for both Zeeman sublevels to try and remain close to two-state behaviour, then $\kappa = 1.002$ \cite{Kempen2002}, and even if we consider using one level from the $F=2$ manifold and one in the $F=1$ manifold and coupling the two via an optical Raman transition, the phase mismatch is at best a few percent \cite{Widera2006}.

\section{Conclusion}
\label{secConclusion}

We have investigated the evolution of a two-state BEC where the two states $|0\rangle$ and $|1\rangle$ are coupled in some way, and the s-wave scattering lengths between $(|0\rangle, |0\rangle)$, $(|1\rangle,|1\rangle)$ and $(|0\rangle,|1\rangle)$ are not all identical. Our analysis demonstrates that the mean field background of such a system exhibits periodicity, in that the same physical state occurs repeatedly with a fixed period. 

To investigate the stability and energy excitation spectrum of the system, we performed a perturbation analysis beyond the mean field. As the mean field background about which we are considering perturbations is time-dependent, the standard approach would be to apply the Bogoliubov-de Gennes theory, which would result in a purely numerical solution to the dynamics. As our system is time-dependent but \emph{periodic}, however, we are able to apply Floquet's theorem to obtain analytic insights into the system.

From the period of the mean field evolution, we can extract the monodromy matrix describing the dynamics, the eigenvalues of which allow us to calculate the Floquet exponents for the system. These Floquet exponents are generally complex, indicating dynamic instabilities within the system. Specifically, the real part of the exponents describe the energy spectrum of excitations within the condensate, while the imaginary part corresponds to exponential growth of the excitations. As the exponents are momentum-dependent, this results in the amplification of excitations at very specific momenta.

The evolution of the annihilation and creation operators of the excitations are identical to those corresponding to the annihilation and creation operators of the photons in optical degenerate parametric down-conversion. Optical parametric down-conversion is a well-studied and often-employed scheme to generate EPR-entangled pairs of photons and, just as in the optical case, our scheme results in EPR entanglement between two particle excitations with the same momentum magnitude but opposite direction.

As the amplification can only occur if there exists substantial condensate population within a few de Broglie wavelengths of the initial excitation, it is possible to create a situation when the entangled excitations are emitted directionally. This can be done by employing a high aspect ratio condensate, where the Thomas-Fermi radius in the tightly trapped axis is comparable to or less than the de Broglie wavelength of the lowest momentum excitation, resulting in exponential gain predominantly along the long axis of the condensate. The narrower the condensate is, the more directional the entangled beams can become.

The system we have analyzed in this paper is not difficult to arrange experimentally, and we have shown that an experiment matching our criteria resulted in particle emission along the long axis of the condensate at momentum that matches our predictions \cite{Dall2009}. That experiment was not optimized for the production of entangled particle beams, however, and, as we have discussed, a better scheme would involve an optically rather than magnetically trapped condensate.

We would like to thank Joseph Hope and Karen Kheruntsyan for helpful discussions and suggestions. This work was funded by the ARC Centre of Excellence for Quantum Atom Optics and ARC Discovery Project scheme. We also acknowledge support from the NCI National Supercomputing Facility.

\appendix
\section{Evolution of the excitations}
\label{appExcitationEvolution}
The theory of elementary excitations in unstable Bose-Einstein condensates has been considered before \cite{Leonhardt2003}.  In this appendix, restrictions on the Floquet exponents for the system under consideration are obtained from which the equations of motion for the instabilities may be solved. This is achieved by applying the methods discussed in \cite{Leonhardt2003} and consideration of the symmetries of the system Hamiltonian.  The excitation dynamics is described by a temporally periodic but spatially homogenous and isotropic Hamiltonian.  This Hamiltonian can be obtained by making a perturbative expansion of the Hamiltonian Eq.~(\ref{eqInitialHamiltonian}) considered in Section~\ref{secPeriodicSolutions} about the time-dependent (but periodic) mean-field.

Consider a general quadratic, spatially-homogeneous Hamiltonian $\hat{H}(t)$ of period $T$ in terms of the operators $\hat{\phi}_i(\mathbf{x}, t)$ which obey the usual equal-time bosonic commutation relations. Due to conservation of momentum, the Hamiltonian $\hat{H}(t)$ when written in the Fourier basis can only contain terms of the form $\hat{\phi}_i^\dagger(\mathbf{k}, t) \hat{\phi}_j^{\phantom{\dagger}}(\mathbf{k}, t)$, $\hat{\phi}_i^\dagger(\mathbf{k}, t) \hat{\phi}_j^{\dagger}(-\mathbf{k}, t)$ and $\hat{\phi}_i^{\phantom{\dagger}}(\mathbf{k}, t) \hat{\phi}_j^{\phantom{\dagger}}(-\mathbf{k}, t)$, while terms of the form $\hat{\phi}_i^\dagger(\mathbf{k}, t) \hat{\phi}_j(-\mathbf{k}, t)$, $\hat{\phi}_i^\dagger(\mathbf{k}, t)\hat{\phi}_j^\dagger(\mathbf{k}, t)$, and $\hat{\phi}_i(\mathbf{k}, t) \hat{\phi}_j(\mathbf{k}, t)$ cannot occur. As the Hamiltonian $\hat{H}(t)$ is homogeneous by assumption, the operator equations of motion will take their simplest form in a Fourier basis. In this basis the equations of motion for the operators $\hat{\phi}_i(\mathbf{k}, t)$ can be written in matrix form
\begin{subequations}
    \label{eqMatrixOperatorEvolution}
    \begin{align}
        i \hbar \frac{\partial }{\partial t}\hat{\bm{\Upsilon}}(\mathbf{k}, t) &= \mathcal{H}(\mathbf{k}, t) \hat{\bm{\Upsilon}}(\mathbf{k}, t), \label{eqUpsilonEquationOfMotion}\\
        \hat{\bm{\Upsilon}}(\mathbf{k}, t) &= 
        \begin{pmatrix}
            \hat{\phi}_1^{\phantom{\dagger}}(\mathbf{k}, t) &
            \hat{\phi}_1^\dagger(-\mathbf{k}, t) &
            \hat{\phi}_2^{\phantom{\dagger}}(\mathbf{k}, t) &
            \hat{\phi}_2^\dagger(-\mathbf{k}, t) &
            \dots
        \end{pmatrix}^\text{T},
    \end{align}
\end{subequations}
where the matrix $\mathcal{H}(\mathbf{k}, t)$ obeys
\begin{align}
        \mathcal{H}(\mathbf{k}, t+T) &= \mathcal{H}(\mathbf{k}, t),\\
        \mathcal{H}(\mathbf{k}, t) &= \mathcal{H}(-\mathbf{k}, t), \label{eqHReflectionSymmetry}
\end{align}
where the last equality holds because $\hat{H}(t)$ is isotropic.

If the Hamiltonian $\hat{H}(t)$ were not time-dependent, $\mathcal{H}(\mathbf{k}, t)$ could be diagonalized to find the (potentially complex) eigenvalues $\Omega_j(\mathbf{k})$ and corresponding operators $\hat{Q}_j(\mathbf{k}, t)$ which would evolve as
\begin{align}
    \label{eqContinouousTimeEigenoperators}
    i \hbar \frac{\partial }{\partial t}\hat{Q}_j(\mathbf{k}, t) &= \hbar \Omega_j(\mathbf{k}) \hat{Q}_j(\mathbf{k}, t),
\end{align}
where the $\hat{Q}_j(\mathbf{k}, t)$ need not obey boson commutation relations. The real parts of the eigenvalues of $\mathcal{H}(\mathbf{k}, t)$ would give the excitation spectrum of the Hamiltonian $\hat{H}(t)$ with non-zero imaginary components giving the growth rates of the corresponding unstable mode.

In the case of a periodic matrix $\mathcal{H}(\mathbf{k}, t)$ it is instead the monodromy matrix $\mathcal{M}(\mathbf{k})$ (see Section~\ref{secInstabilitiesAndExcitations}) that we wish to diagonalize. The monodromy matrix $\mathcal{M}(\mathbf{k})$ satisfies
\begin{align}
    \label{eqMonodromyMatrixAppendix}
    \hat{\mathbf{\Upsilon}}(\mathbf{k}, nT) &= \mathcal{M}(\mathbf{k})^n \hat{\bm{\Upsilon}}(\mathbf{k}, 0),
\end{align}
where $n$ is a positive integer. In place of Eq.~(\ref{eqContinouousTimeEigenoperators}) we seek the operators $\hat{Q}_j(\mathbf{k}, t)$ that obey
\begin{align}
    \label{eqQOperatorEvolution}
    \hat{Q}_j(\mathbf{k}, T) &= \lambda_j(\mathbf{k}) \hat{Q}_j(\mathbf{k}, 0),
\end{align}
where the $\hat{Q}_j(\mathbf{k}, t)$ are defined by
\begin{align}
    \label{eqQOperatorDefinition}
    \hat{Q}_j(\mathbf{k}, t) &= \bm{c}_j^\dagger(\mathbf{k}) \hat{\bm{\Upsilon}}(\mathbf{k}, t),
\end{align}
for some vectors $\bm{c}_j(\mathbf{k})$, where $\bm{c}_j^\dagger(\mathbf{k})$ denotes the conjugate transpose. 

Using definitions Eq.~(\ref{eqMonodromyMatrixAppendix})--Eq.~(\ref{eqQOperatorDefinition}), it follows that the $\lambda_j(\mathbf{k})$ and $\bm{c}_j^\dagger(\mathbf{k})$ are respectively the eigenvalues and left eigenvectors of $\mathcal{M}(\mathbf{k})$,
\begin{align}
    \hat{Q}_j(\mathbf{k}, T) = \bm{c}_j^\dagger(\mathbf{k}) \hat{\bm{\Upsilon}}(\mathbf{k}, T) &= \bm{c}_j^\dagger(\mathbf{k}) \mathcal{M}(\mathbf{k}) \hat{\bm{\Upsilon}}(\mathbf{k}, 0) \\
    \hat{Q}_j(\mathbf{k}, T) = \lambda_j(\mathbf{k}) \hat{Q}_j(\mathbf{k}, 0) &= \lambda_j(\mathbf{k}) \bm{c}_j^\dagger \hat{\bm{\Upsilon}}(\mathbf{k}, 0)\\
    \implies \bm{c}_j^\dagger(\mathbf{k}) \mathcal{M}(\mathbf{k}) \hat{\bm{\Upsilon}}(\mathbf{k}, 0) &= \lambda_j(\mathbf{k}) \bm{c}_j^\dagger \hat{\bm{\Upsilon}}(\mathbf{k}, 0) \label{eqLeftEigenvectorWithOperator}\\
    \implies \bm{c}_j^\dagger(\mathbf{k}) \mathcal{M}(\mathbf{k}) &= \lambda_j(\mathbf{k}) \bm{c}_j^\dagger(\mathbf{k}), \label{eqLeftEigenvector}
\end{align}
where Eq.~(\ref{eqLeftEigenvector}) follows from Eq.~(\ref{eqLeftEigenvectorWithOperator}) as the components of $\hat{\bm{\Upsilon}}(\mathbf{k}, 0)$ are linearly independent operators.

The operators $\hat{Q}_j(\mathbf{k}, t)$ are not necessarily bosonic annihilation or creation operators. To determine the conditions under which they are, we consider their Hermitian conjugates $\hat{Q}_j^\dagger(\mathbf{k}, t)$. As every operator in $\hat{\bm{\Upsilon}}(-\mathbf{k}, t)$ is the Hermitian conjugate of an operator in $\hat{\bm{\Upsilon}}(\mathbf{k}, t)$, the $\hat{Q}_j^\dagger(\mathbf{k}, t)$ can be written as
\begin{align}
    \label{eqQDaggerDefinition}
    \hat{Q}_j^\dagger(\mathbf{k}, t) &= \bm{d}_j^\dagger(\mathbf{k}) \hat{\bm{\Upsilon}}(-\mathbf{k}, t)
\end{align}
for some vectors $\bm{d}_j(\mathbf{k})$. It follows from Eq.~(\ref{eqQOperatorEvolution}) that the $\hat{Q}_j^\dagger(\mathbf{k}, t)$ will obey
\begin{align}
    \label{eqQDaggerEvolution}
    \hat{Q}_j^\dagger(\mathbf{k}, T) &= \lambda_j^*(\mathbf{k}) \hat{Q}_j^\dagger(\mathbf{k}, 0).
\end{align}

The commutators of the $\hat{Q}_j^{(\dagger)}(\mathbf{k}, t)$ will be constant as they are constant linear combinations of the $\hat{\phi}^{(\dagger)}(\pm\mathbf{k}, t)$, the commutators of which are themselves constant. Using this requirement gives
\begin{align}
    \left[ \hat{Q}_i^{\phantom{\dagger}}(\mathbf{k}, T),\, \hat{Q}_j^\dagger(\mathbf{k}, T) \right] &= \left[ \lambda_i(\mathbf{k}) \hat{Q}_i^{\phantom{\dagger}}(\mathbf{k}, 0),\, \lambda_j^*(\mathbf{k}) \hat{Q}_j^\dagger(\mathbf{k}, 0)\right]\\
        &= \lambda_i(\mathbf{k}) \lambda_j^*(\mathbf{k}) \left[ \hat{Q}_i^{\phantom{\dagger}}(\mathbf{k}, 0),\, \hat{Q}_j^\dagger(\mathbf{k}, 0)\right].
        \label{eqInvariantCommutator}
\end{align}
For Eq.~(\ref{eqInvariantCommutator}) to be true either $\lambda_i(\mathbf{k}) \lambda_j^*(\mathbf{k}) = 1$ or the two operators commute. Specifically, for $\hat{Q}_i(\mathbf{k}, t)$ to be an annihilation or creation operator it is required that ${\lambda_i^*(\mathbf{k})}^{-1} = \lambda_i(\mathbf{k})$. In terms of the Floquet exponents (see Section~\ref{secInstabilitiesAndExcitations}) $\displaystyle \xi_i(\mathbf{k}) = \frac{i}{T} \ln \lambda_i(\mathbf{k})$, this requirement becomes $\xi_i(\mathbf{k}) = \omega_i(\mathbf{k})$. Hence it is only for purely real Floquet exponents that the eigenvalues of $\mathcal{M}(\mathbf{k})$ correspond to bosonic annihilation or creation operators. Note that in the degenerate case in which $\mathcal{H}(\mathbf{k}, t)$ is time-independent, the Floquet exponents $\xi(\mathbf{k})$ are equal to the eigenvalues $\Omega(\mathbf{k})$ of $\mathcal{H}(\mathbf{k}, t)$. Hence the real components of the Floquet exponents are related to the excitation spectrum and non-zero imaginary components are related to the existence of instabilities.

Generally the Floquet exponents $\xi_i$ may have a non-zero imaginary component. In this case the $\hat{Q}_i(\mathbf{k}, t)$ will not be bosonic annihilation or creation operators, although such operators can be constructed from linear combinations of the $\hat{Q}_i(\pm\mathbf{k}, t)$. Before constructing such operators, we first consider the restrictions on the possible eigenvalues of $\mathcal{M}(\mathbf{k})$.

First it is noted that if $\lambda_i(\mathbf{k})$ is an eigenvalue of $\mathcal{M}(\mathbf{k})$, then from Eq.~(\ref{eqQDaggerDefinition})--Eq.~(\ref{eqQDaggerEvolution}) $\lambda_i^*(\mathbf{k})$ must be an eigenvalue of $\mathcal{M}(-\mathbf{k})$. However, from the reflection symmetry of $\mathcal{H}(\mathbf{k})$ defined by Eq.~(\ref{eqHReflectionSymmetry}) we have that $\mathcal{M}(-\mathbf{k}) = \mathcal{M}(\mathbf{k})$ and hence $\lambda_i^*(\mathbf{k})$ must also be an eigenvalue of $\mathcal{M}(\mathbf{k})$.

Secondly, not all of the operators $\hat{Q}_i^{(\dagger)}(\mathbf{k}, t)$ can commute. As the $\hat{Q}_i^{(\dagger)}(\mathbf{k}, t)$ form a complete basis over the same space as the $\hat{\phi}_i^{(\dagger)}(\mathbf{k}, t)$ which themselves do not all commute, for every operator $\hat{Q}_i(\mathbf{k}, t)$ there must be at least one other operator with which it does not commute. From Eq.~(\ref{eqInvariantCommutator}) then follows the requirement that if $\lambda_i(\mathbf{k})$ is an eigenvalue,   ${\lambda_i^*(\mathbf{k})}^{-1}$ must also be an eigenvalue.

Combining these two requirements gives a consistency condition for the eigenvalues of $\mathcal{M}(\mathbf{k})$: if $\lambda$ is an eigenvalue, $\lambda^*$, $\lambda^{-1}$, and ${\lambda^*}^{-1}$ must all be eigenvalues. These conditions can be met using 1, 2 or 4 distinct eigenvalues of $\mathcal{M}(\mathbf{k})$.

For the degenerate case in which all of $\lambda$, $\lambda^*$, $\lambda^{-1}$, and ${\lambda^*}^{-1}$ are equal, the eigenvalue $\lambda = \pm 1$. The corresponding Floquet exponent is $\xi = 0$ or $\xi = \pi\nu_0$ where $\displaystyle \nu_0 = T^{-1}$. This is not an interesting case and does not occur in Figure~\ref{figCondensateEigenvalues}.

There are two ways that two distinct eigenvalues can be used to satisfy the consistency condition for the eigenvalues. The first possibility is that $\lambda = {\lambda^*}^{-1}$ (with $\lambda^*$ being the second eigenvalue). In this case the Floquet exponents are $\xi = \pm \omega$. In this case, the operators corresponding to the eigenvalues are bosonic annihilation or creation operators as shown above. The second possibility is that $\lambda = \lambda^*$ (with $\lambda^{-1}$ being the second eigenvalue). In this case the Floquet exponents are $\xi = \pm i\gamma$ or $\xi = \pi \nu_0 \pm i\gamma$. This situation is seen in Figure~\ref{figCondensateEigenvalues} around $k \approx 1.5 \times 10^6$m$^{-1}$ and $k \approx 2.25 \times 10^6$m$^{-1}$.

The final possibility is that four distinct eigenvalues are used to satisfy the consistency condition. In this case the four eigenvalues $\lambda$, $\lambda^*$, $\lambda^{-1}$, and ${\lambda^*}^{-1}$ are different and the corresponding Floquet exponents are $\xi = \omega \pm i\gamma$ and $\xi' = -\omega \pm i\gamma$. It is this situation that occurs in Figure~\ref{figCondensateEigenvalues} around $k \approx 0.75 \times 10^6$m$^{-1}$.

In summary, the Floquet exponents with non-zero real parts come in pairs $\xi(\mathbf{k}) = \omega(\mathbf{k}) \pm i\gamma(\mathbf{k})$. From the operators corresponding to these pairs of exponents bosonic annihilation and creation operators can be constructed.

Consider the eigenvalues $\displaystyle \lambda = e^{r + i \phi}$ and $\displaystyle \lambda' = e^{-r + i\phi}$, and the corresponding operators $\hat{Q}(\mathbf{k}, t)$ and $\hat{Q}'(\mathbf{k}, t)$. From these operators we define the following two operators which will respectively be shown to be bosonic annihilation and creation operators,
\begin{align}
    \hat{\Lambda}_1(\mathbf{k}, t) &= \frac{1}{\sqrt{2}} \left( \hat{Q}(\mathbf{k}, t) + \hat{Q}'(\mathbf{k}, t)\right),\\
    \hat{\Lambda}_2(\mathbf{k}, t) &= \frac{1}{\sqrt{2}} \left( \hat{Q}(\mathbf{k}, t) - \hat{Q}'(\mathbf{k}, t)\right).
\end{align}
As $\lambda \lambda'^* = 1$, $\hat{Q}(\mathbf{k}, t)$ and $\hat{Q}'^\dagger(\mathbf{k}, t)$ will not commute. By appropriate rescaling of the operators, we can define their commutator to be
\begin{align}
    \left[ \hat{Q}(\mathbf{k}, t),\, \hat{Q}'^\dagger(\mathbf{k}, t) \right] &= 1.
\end{align}
This choice defines the value of the other non-zero commutator,
\begin{align}
    \left[ \hat{Q}'(\mathbf{k}, t),\, \hat{Q}^\dagger(\mathbf{k}, t) \right] &= 1.
\end{align}

From these two commutators it can then be shown that $\hat{\Lambda}_1(\mathbf{k}, t)$ obeys the commutation relations appropriate for an annihilation operator, while $\hat{\Lambda}_2(\mathbf{k}, t)$ obeys the commutation relations for a creation operator. For example,
\begin{widetext}
\begin{align}
    \left[\hat{\Lambda}_1^{\phantom{\dagger}}(\mathbf{k}, t),\, \hat{\Lambda}_1^\dagger(\mathbf{k}, t) \right] &= \frac{1}{2} \left[ \hat{Q}(\mathbf{k}, t) + \hat{Q}'(\mathbf{k}, t),\, \hat{Q}^\dagger(\mathbf{k}, t) + \hat{Q}'^\dagger(\mathbf{k}, t)\right] = 1.
\end{align}

Defining $\hat{\Lambda}_1'(-\mathbf{k}, t) = \hat{\Lambda}_2^\dagger(\mathbf{k}, t)$, the evolution of the operators $\hat{\Lambda}_1(\mathbf{k}, t)$ and $\hat{\Lambda}_1'(\mathbf{k}, t)$ can now be determined. $\hat{\Lambda}_1(\mathbf{k}, t)$ evolves as
\begin{align}
    \hat{\Lambda}_1(\mathbf{k}, nT) &= \frac{1}{\sqrt{2}} \left(\hat{Q}(\mathbf{k}, nT) + \hat{Q}'(\mathbf{k}, nT) \right) \\
        &= \frac{1}{\sqrt{2}} \left( e^{nr + i n\phi} \hat{Q}(\mathbf{k}, 0) + e^{-nr + i n \phi}\hat{Q}'(\mathbf{k}, 0)\right) \\
        \begin{split}
            &=  e^{i n \phi} \bigg[ \frac{1}{2}\left( e^{nr} - e^{-nr}\right)\frac{1}{\sqrt{2}}\left(\hat{Q}(\mathbf{k}, 0) - \hat{Q}'(\mathbf{k}, 0) \right) + \frac{1}{2} \left( e^{nr} + e^{-nr}\right)\frac{1}{\sqrt{2}} \left( \hat{Q}(\mathbf{k}, 0) + \hat{Q}'(\mathbf{k}, 0)\right)\bigg]
        \end{split}\\
        &= e^{i n \omega T} \left( \sinh(n\gamma T) \hat{\Lambda}_1'^\dagger(-\mathbf{k}, 0) + \cosh(n \gamma T) \hat{\Lambda}_1(\mathbf{k}, 0)\right), \label{eqLambdaOperatorEvolution}
\end{align}
\end{widetext}
where $n$ is a positive integer. Similarly, $\hat{\Lambda}_1'(\mathbf{k}, t)$ can be shown to evolve as
\begin{align}
    \hat{\Lambda}_1'(\mathbf{k}, nT) = e^{-i n \omega T} \bigg( & \sinh(n \gamma T) \hat{\Lambda}_1^\dagger(-\mathbf{k}, 0) \nonumber \\
& + \cosh(n \gamma T) \hat{\Lambda}_1'(\mathbf{k}, 0) \bigg).
\label{eqLambdaPrimeOperatorEvolution}
\end{align}

Finally, if the $\hat{Q}(\mathbf{k}, t)$ and $\hat{Q}'(\mathbf{k}, t)$ operators correspond to only two distinct eigenvalues (i.e.\ $\lambda = \lambda^*$), then by the uniqueness of the eigenvectors $\hat{\Lambda}_1'(\mathbf{k}, t) = \hat{\Lambda}_1(\mathbf{k}, t)$. For this case $\displaystyle e^{i n\omega T} = \pm 1$, making Eq.~(\ref{eqLambdaOperatorEvolution}) and Eq.~(\ref{eqLambdaPrimeOperatorEvolution}) consistent.

\section{Proof of periodicity of the nonlinear Bloch equations}
\label{appNoLimitCyles}

In this Appendix we give a proof that the solutions to the nonlinear optical Bloch equations considered in Section~\ref{secPeriodicSolutions}, which describe the dynamics of the mean field of a two-level homogeneous condensate in the case that the scattering lengths for the two levels are not equal, are periodic. We restate the nonlinear optical Bloch equations (\ref{eqOpticalBlochEquations}) here for convenience:
\begin{subequations}
    \label{eqAppendixOpticalBlochEquations}
    \begin{align}
        \frac{d}{dt}\rho_{10} &= -i\frac{g}{2} (1-w)\rho_{10} + i \Omega w,\\
        \frac{d }{dt}w &= -4 \Omega {\mathrm{Im}} \{\rho_{10}\},
    \end{align}
\end{subequations}
where $\rho_{10}$ is the off-diagonal element of the density matrix, and $w = \rho_{11} - \rho_{00}$.

These equations describe motion on the surface of a sphere, known as the Bloch sphere (see Figure~\ref{figBlochSphere}), with each point on the sphere corresponding to a different physical state.  A generalisation of the Poincar{\'{e}}-Bendixson theorem that applies to compact, connected, two-dimensional orientable manifolds (such as the surface of the sphere) states that every trajectory either approaches one or more fixed points, or approaches a periodic orbit \cite{Schwartz1963}.  The possibility of a space-filling trajectory is precluded as the surface of a sphere is not homeomorphic to a torus (see \cite{Schwartz1963}).  Periodic trajectories and fixed points trivially approach themselves.

We next show that there are no limit cycles in this system, i.e.\ there exist no trajectories that approach periodic orbits which are not themselves periodic.

We first write the equations of motion for the system in the usual spherical polar coordinates as
\begin{subequations}
    \label{eqAppendixEvolutionOfAngles}
    \begin{align}
        \frac{d \theta}{dt} &= 2 \Omega \sin\phi, \label{eqAppendixEvolutionOfAnglesTheta}\\
        \frac{d \phi}{dt} &= 2 \Omega \cos\phi \cot\theta - \frac{1}{2} g (1-\cos\theta), \label{eqAppendixEvolutionOfAnglesPhi}
    \end{align}
\end{subequations}
where we note that these equations are of the form
\begin{align}
    \dot{\mathbf{r}} &= \frac{2}{\hbar} \hat{\mathbf{r}} \times \nabla E, \label{eqAppendixConservativeEvolution}
\end{align}
where $E$ is the energy per particle previously defined in Eq.~(\ref{eqOpticalBlochEnergy}), and a hat is used to denote unit vectors.

Eq.~(\ref{eqAppendixConservativeEvolution}) shows that a given point moves in the direction perpendicular to the gradient of $E$ at a rate proportional to the magnitude of the gradient, and hence the energy $E$ is conserved along any given trajectory.

Assuming a limit cycle exists, there will exist a trajectory approaching the limit cycle which is not the limit cycle. For each point in the limit cycle $\mathbf{x}$ there exists an infinite sequence of points $\mathbf{s}_i$ that approach $\mathbf{x}$ which are the intersection of a curve which intersects no trajectory tangentially (a \emph{transversal}, see \cite{Schwartz1963}) and the trajectory approaching the limit cycle.  As the energy $E$ is continuous, we have $\displaystyle \lim_{i\rightarrow\infty}E(\mathbf{s}_i) = E(\mathbf{x})$, and therefore the energy of the approaching trajectory must be the same as the energy of the limit cycle. Further, as energy is conserved along the trajectory, and all the points $\mathbf{s}_i$ lie on a trajectory, $\displaystyle E(\mathbf{s}_i) = E(\mathbf{x})$ for {\textit{all}} points $\mathbf{s}_i$. Thus the derivative of $E$ in the direction of the transversal is zero as $\displaystyle \lim_{i\rightarrow\infty} \frac{E(\mathbf{s}_i) - E(\mathbf{x})}{| \mathbf{s}_i - \mathbf{x} | } = 0$.  It is already known that the derivative of the energy is zero parallel to the limit cycle as energy is conserved along any trajectory. As the direction of the transversal and the limit cycle at $\mathbf{x}$ are linearly independent (a transversal intersects no trajectory tangentially) and the manifold is two-dimensional, the gradient of $E$ at $\mathbf{x}$ must necessarily be zero.  From (\ref{eqAppendixConservativeEvolution}) this implies that $\mathbf{x}$ is a fixed point, which is a contradiction with $\mathbf{x}$ being part of a limit cycle.  Our original assumption that a limit cycle exists is therefore false.

It now remains to show that no trajectories approach fixed points (with the exception of the fixed points themselves).  As the possibility of limit cycles has been excluded, the generalised Poincar{\'{e}}-Bendixson theorem shows that all remaining trajectories are periodic.

To analyse the stability of the fixed points, it is first necessary to determine their location.  It immediately follows from (\ref{eqAppendixEvolutionOfAnglesTheta}) that fixed points will satisfy $\sin\phi = 0$, i.e. fixed points are restricted to the great circle in the $x$--$z$ plane.  To parameterise this circle by a single coordinate, we alter the usual ranges of the spherical polar coordinates such that $\phi \in \left[-\frac{\pi}{2}, \frac{\pi}{2}\right)$ and $\theta \in [-\pi, \pi)$.  With this definition $\cos\phi = 1$ over the entire circle, instead of $\cos\phi = \pm 1$ over opposite halves.

Using the tan-half substitution $s = \tan\frac{1}{2}\theta$, it can be shown from (\ref{eqAppendixEvolutionOfAnglesPhi}) that all fixed points satisfy
\begin{align}
    g s^3 &= \Omega (1 - s^4). 
    \label{eqAppendixFixedPointsEquation}
\end{align}

The stability of (\ref{eqAppendixEvolutionOfAngles}) about the fixed points may be used to demonstrate that there are no trajectories that approach the fixed points (with the exception of the fixed points themselves).  Linearizing (\ref{eqAppendixEvolutionOfAngles}) about the fixed points yields
\begin{align}
    \frac{d}{dt} \delta \theta &= 2 \Omega \delta \phi,\\
    \frac{d}{dt} \delta \phi &= - \frac{\Omega}{2 s^2}(s^4 + 3) \delta \theta,
\end{align}
where $\delta \theta$ and $\delta\phi$ are the deviations from the fixed point and the identity \eqref{eqAppendixFixedPointsEquation} has been used to simplify the result.  The eigenvalues of this linear system of differential equations are
\begin{align}
    \lambda &= \pm\sqrt{- \frac{\Omega^2}{s^2}(s^4 + 3)},
\end{align}
which are pure imaginary as both $\Omega$ and $s$ are real.  This implies that there exist no trajectories near any fixed points of the system that asymptotically approach the fixed point.  It may then be concluded that with the exception of the fixed points themselves, all trajectories of the system (\ref{eqAppendixOpticalBlochEquations}) are periodic, and Floquet's theorem may be applied in Section~\ref{secInstabilitiesAndExcitations}.

\end{document}